\documentclass[preprint,3p,times]{elsarticle}
\pdfoutput=1 
\makeatletter
\def\ps@pprintTitle{%
 \let\@oddhead\@empty
 \let\@evenhead\@empty
 \def\@oddfoot{}%
 \let\@evenfoot\@oddfoot}
\makeatother
\usepackage{multirow}
\usepackage[caption=false,subrefformat=parens,labelformat=parens]{subfig}
\usepackage{natbib}
\usepackage[]{units}
\usepackage[usenames,dvipsnames]{color}
\usepackage{xcolor}
\usepackage[percent]{overpic}
\usepackage{mathtools}
\usepackage{tikz}
\usetikzlibrary{plotmarks}

\usepackage{multirow}
\newcommand\crule[3][black]{\textcolor{#1}{\rule{#2}{#3}}}

\definecolor{myblue}{RGB}{0, 0, 232}
\definecolor{mygreen}{RGB}{0, 132, 0}
\definecolor{myred}{RGB}{222, 0, 0}
\definecolor{mygreen2}{RGB}{0, 153, 76}
\definecolor{myblue2}{RGB}{102, 102, 255}

\newcommand{\myCirc}{\begin{tikzpicture}\node[mark size=1.8pt,color=myblue] at (0,0) {\pgfuseplotmark{*}}; \end{tikzpicture}}
\newcommand{\mySqua}{\begin{tikzpicture}\node[mark size=1.8pt,color=mygreen] at (0,0) {\pgfuseplotmark{square*}}; \end{tikzpicture}}
\newcommand{\myTri}{\begin{tikzpicture}\node[mark size=2pt,color=myred] at (0,0) {\pgfuseplotmark{triangle*}}; \end{tikzpicture}}

\newcommand{\vcol}{v_\text{imp}}
\newcommand{\dtil}{\widetilde{\delta}}
\newcommand{\dtilmax}{\widetilde{\delta}^\text{max}}

\newcommand{\atilmax}{\widetilde{a}^\text{max}}
\newcommand{\Ftil}{\widetilde{F}_N}

\begin{document}
	\begin{frontmatter}

	\title{The effect of surface geometry on collisions between nanoparticles}

	\author[buf,oist]{Yoichi Takato\corref{cor1}}
	\ead{ytakato@buffalo.edu}

	\author[buf]{Michael E. Benson}
	\ead{mebenson@buffalo.edu}

	\author[buf]{Surajit Sen\corref{cor2}}
	\ead{sen@buffalo.edu}

	\address[buf]{Department of Physics \\ The State University of New York at Buffalo, Buffalo, New York 14260-1500, USA}
	\address[oist]{Mathematical Soft Matter Unit \\ Okinawa Institute of Science and Technology Graduate University, Onna-son, Okinawa, 904-0495 Japan}
	\cortext[cor1]{Corresponding author}
	\cortext[cor2]{Principal corresponding author}

	\date{\today}

	\begin{abstract}
In this molecular dynamics study, we examine the local surface geometric
effects of the normal impact force between two approximately
spherical nanoparticles that collide in a vacuum. Three types of surface
geometries, facets, sharp crystal edges, and amorphous surfaces of
nanoparticles with radii \(R < \unit[10]{nm}\) are considered, and the
impact force is compared with its macroscopic counterpart described by a
nonlinear contact force, \(F_{N} \propto \delta^{n}\) with \(n = 3/2\)
derived by Hertz (1881), where \(\delta\) is the overlap induced by
elastic compression. We study the surface geometry-dependent impact
force. For facet-facet impact, the mutual contact surface area does not
expand due to the large facet surface, and this in turn leads to a
non-Hertz impact force, \(n < 3/2\). A Hertz-like contact force,
\(n \approx 3/2\), is recovered in the edge contact and in the amorphous
surface contact, allowing expansion of the mutual contact surface area.
The results suggest that collisions of amorphous nanoparticles or nanoparticles with sharp edges may maintain dynamic phenomena, such as
breathers and solitary waves, originating from the nonlinear contact
force.
\end{abstract}


\end{frontmatter}

\section{Introduction}\label{introduction}

The discrete nature of nanoscale materials has often revealed surprising
phenomena. Nanoscale normal contact force and friction laws depend strongly on
the contact. A molecular dynamics (MD) study by Luan and Robbins
\cite{luan2005} demonstrated the breakdown of continuum theory in a system of
nanoscale non-adhesive cylinders made to model contact between substrates
and atomic force microscope tips. The normal contact force for rough
surfaces of solids with amorphous and crystal structures revealed
significant departures from the Hertz contact force for contacting
elastic perfect spheres \cite{hertz1881}. Atomically rough surfaces on contacting objects
influence the contact considerably, and local surface geometry in the
vicinity of the contact region promotes variations of the contact
surface area \cite{luan2005,pastewka2016}.

Atomic surface asperities also dictate impact phenomena in
nanoparticles. Gold nanoparticles can take a near-spherical shape, one
of the thermodynamically stable shapes \cite{barnard2009}, and the surface of such a nanoparticle is comprised of
crystal steps and terraces. Facets and crystal steps in small
nanoparticles can make them ``rough'', which in turn affects the nature
of their interactions. The lack of spherical nanoparticle symmetry often
leads to more complex dynamics than the Hertz contact theory for perfect
spheres. Roll, slide, and deflection of colliding nanoparticles arising
from shape asymmetry have been observed elsewhere \cite{dominik1996,awasthi2007}.

In dynamics, the coefficient of restitution often suffices
to predict overall behavior of a dissipative many-body system such as clustering of dissipative particles \cite{luding1999}. Precise measurement
of impact force, however, is critical for certain dynamic systems. A
recent MD simulation demonstrated that a one-dimensional
chain of nanoscale buckyballs can permit the propagation of a solitary
wave \cite{xu2016}, which is a non-dispersive propagating wave in a macroscopic granular system
discovered by Nesterenko \cite{nesterenko1983}. Realization of the solitary wave at nanoscale
implies that the interaction between buckyballs is described by a
nonlinear force. That is, the power of overlap must satisfy \(n > 1\) so
that a sharp propagating pulse of energy can be formed \cite{nesterenko2001,sen2008}.

Direct observation of the impact force of repulsive nanoparticles made by MD simulations, on the other hand, has been reported
in a few papers \cite{kuninaka2009a,kim2010jkps,zeng2010}. The nature of impact forces between rough surfaces of
colliding nanoparticles is not fully understood yet. Our MD
study presents precise details of contact forces for collisions between nanoparticles
having three different contact surface geometries, in order to
investigate the influence of surface roughness in a systematic fashion.
The surface of an amorphous nanoparticle and two surface geometries,
facets and sharp crystal edges of a face-centered cubic (fcc) crystal nanoparticle, are considered as contact surfaces.

Many MD studies on nanoparticle collisions indicate that
nanoparticles are highly elastic if the impact velocity is kept below
their material yield point, although a small amount of the initial
kinetic energy admittedly dissipates during the impact process  \cite{kuninaka2009a,kuninaka2009b,han2010,takato2014}. In
addition, a recent study showed that the effect of nanoparticle adhesive
property is negligibly small \cite{takato2015} if the adhesive nanoparticles collide beyond a critical velocity determined by the balance between adhesion and elastic energies \cite{thornton1998,awasthi2006,awasthi2007,jung2010}. Therefore, for simplicity, we consider only repulsive
nanoparticles in our simulation, adopting the Weeks-Chandler-Andersen
(WCA) potential \cite{weeks1971,luan2005,takato2014}, which is a type of Lennard-Jones (LJ) potential modified
to yield a purely repulsive interaction.

In our work, a non-Hertzian contact force exhibiting velocity dependence
is found in monocrystalline nanoparticles impacting on facets. This type of
impact makes the dynamics of a collision considerably different from
Hertzian contact mechanics. For instance, the mutual contact surface
created between two nanoparticles in contact remains unchanged, as
opposed to the evolving contact surface area of a Hertzian sphere. In
contrast, the Hertzian contact force is a good approximation for
crystalline nanoparticles impacting on sharp crystal step edges.
Furthermore, amorphous nanoparticles with relatively smooth surfaces and
substantially spherical shapes yield an impact force that resembles the
Hertzian contact force.

The present work is organized as follows: the Hertz contact theory is
briefly reviewed in Sec.~\ref{hertz-contact-theory}. Sec.~\ref{numerical-simulations} discusses our nanoparticle
models and computational methods. Contact forces for
colliding nanoparticles made from amorphous and crystal structures are
displayed in Sec.~\ref{impact-force}. Furthermore, the maximum contact area and maximum
compression for facet contact for monocrystalline nanoparticles are compared
with those for Hertzian spheres in Sec.~\ref{dynamic-behaviors-for-facet-contact}. In Sec.~\ref{discussion}, a departure is discussed. The conclusions are presented in Sec.~\ref{conclusions}.

\section{Hertz contact theory}\label{hertz-contact-theory}

Hertz derived a normal compressive force \(F_\text{H}\) between two
statically contacting elastic spheres that have smooth surfaces \cite{hertz1881,johnson1987}. The
force is expressed in terms of overlap \(\delta = (2R - d)/2\) for two
identical spheres of diameter \(2R\) and center-to-center intersphere
distance \(d\) under compression,
\begin{equation}
F_\text{H} = \kappa_\text{H}\delta^{n}, \label{eq:fh}
\end{equation}
where \(n = 3/2\) and \(\kappa_\text{H} = (4/3)E^{*}R^{*1/2}\). The reduced
Young's modulus is \(E^{*} = E/\lbrack 2(1 - \nu^{2})\rbrack\) with
Young's modulus \(E\) and Poisson ratio \(\nu\). The reduced radius is
\(R^{*} = R/2\) for identical spheres of radius \(R\). The contact force
grows nonlinearly with increasing overlap. The underlying mechanism that
yields the nonlinearity is the varying mutual contact surface area
between the spheres as a function of compression. The shape of the contact surface in the theory
is assumed to be a circle with radius \(a_\text{H}\), from which its area is
\(\pi a_\text{H}^{2}\). General
contact surface shapes and their resultant contact forces including Hertzian force are discussed in Ref.~\cite{sun2011}. The geometrical relation between the displacement and
the contact area under compression gives
\begin{equation}
a_\text{H} = \sqrt{R^{*}\delta}.
\end{equation}
Hence, the contact area expands in proportion to the square root
\(\sqrt{\delta}\) of the overlap.

The equations are derived under static compression. For colliding elastic
spheres that are assumed to follow the Hertz contact theory by ignoring
energy dissipation via surface vibrations, there are several important
quantities that characterize the spheres. A maximum overlap
\(\delta_\text{H}^{\max}\) is the overlap when the colliding spheres
instantaneously come to rest during a collision. Maximum overlap takes
the highest value when the initial kinetic energy associated with the
relative velocity \(v_{\text{imp}}\) is fully converted into elastic
energy to deform the spheres. The maximum overlap \(\delta_\text{H}^{\max}\)
is therefore expressed by Eq.~(3), taking into account the energy
balance between kinetic energy and elastic energy.
\begin{equation}
\delta_\text{H}^{\max} = \left( \frac{15}{16}\frac{M^{*}}{E^{*}R^{*1/2}}{v_{\text{imp}}}^{2} \right)^{2/5}.
\end{equation}

The contact radius for the spheres at maximum compression is then
obtained from Eqs.~(2) and (3),
\begin{equation}
a_\text{H}^{\max} = \sqrt{R^{*}\delta_\text{H}^{\max}} = \left( \frac{15}{16}\frac{M^{*}R^{*2}}{E^{*}}{v_{\text{imp}}}^{2} \right)^{1/5}.
\end{equation}

This reveals that the maximum contact radius scales with impact velocity
to the \(2/5\)th power.

\section{Numerical simulations}\label{numerical-simulations}

\subsection{Models}\label{a.-models}

For an investigation of the dynamic contact force between colliding
nanoparticles, nearly spherical argon nanoparticles \(\xi\) and \(\eta\)
of equal radius \(R\) are prepared. To study the influence of surface
roughness, we employ crystal and amorphous structures as base materials
for making our nanoparticles.

Crystalline nanoparticles are carved out of an fcc
single crystal of solid argon. The resultant nanoparticles have atomic
roughness such as crystal facets and steps on their exterior surfaces
due to their crystal structures (See the insets of Figs.~\subref*{fig:F_edge:2093}--\subref{fig:F_facet:44403}.). We take advantage of the presence of
the surface roughness to obtain contact forces at particular points on
the surfaces. The \{001\} facets are chosen for studying \emph{facet
collisions} and some sharp crystal edges are chosen for studying the
\emph{edge collision} problem.

The amorphous structure is obtained by quenching a molten argon block
from temperature \(T = \unit[70]{K}\) to \(\unit[0.02]{K}\) at a rate
\(\unit[8 \times 10^{10}]{K/s}\) \cite{kristensen1976,nose1985}, followed by equilibrations at \(T_\text{eq} = \unit[0.02]{K}\). A nearly spherical nanoparticle is
then created by cutting the block (See the insets of Figs.~\subref*{fig:F_amorphous:2093}--\subref{fig:F_amorphous:44403}). Radial distribution functions computed
from our equilibrated nanoparticles confirm the amorphous structure
\cite{rahman1976}.

A pure repulsion between two contacting nanoparticles is considered by
employing the WCA potential in Eq.~\eqref{eq:wca} as used for our previous study \cite{takato2014}
and for the Luan and Robbins's study \cite{luan2005}.

\subsection{Interatomic potentials}\label{b.-interatomic-potentials}

Our nonequilibrium MD simulations here use two
interatomic potentials for argon nanoparticles. The shifted 12-6
LJ potential \(V^{\text{LJ}}\) in Eq.~(5) describes an
interatomic interaction between a pair of atoms \(i\) and \(j\) in
either nanoparticle \(\xi\) or nanoparticle \(\eta\), i.e.,
\(i,j \in \xi\) or \(i,j \in \eta\) as follows,
\begin{equation}
V^{\text{LJ}} = \left\{ 4\epsilon\left\lbrack \left( \frac{\sigma}{r_{ij}} \right)^{12} - \left( \frac{\sigma}{r_{ij}} \right)^{6} \right\rbrack + V(r_{c}^{\text{LJ}}) \right\} H(r_{c}^{\text{LJ}} - r_{ij}). \label{eq:lj}
\end{equation}
The \(r_{ij}\) denotes an interatomic distance between $i$-th
and $j$-th atoms, \(\sigma\) is the distance at which the potential is zero, and \(\epsilon\) is the
depth of the potential. The Heaviside step function \(H\) describes
a cutoff of the potential at \(r_{c}^{\text{LJ}} = \unit[2.5]{\sigma}\). The
potential is shifted by \(V(r_{c}^{\text{LJ}})\) to get rid of a
discontinuity at \(r_{ij} = r_{c}^\text{LJ}\) that stems from the adoption
of the cutoff.

In addition, the WCA potential
\(V^{\text{WCA}}\) in Eq.~(6) is adopted to attain purely repulsive
nanoparticles by setting its cutoff radius at
\(r_{c}^{\text{WCA}} = 2^{1/6}\sigma\) where the potential is minimal.
The use of this cutoff value makes the potential purely repulsive
\cite{weeks1971}. This potential applies only to a pair of atoms \(p\) and
\(q\) belonging to nanoparticle \(\xi\) or nanoparticle \(\eta\),
respectively,
\begin{equation}
V^{\text{WCA}} = \left\{ 4\epsilon\left\lbrack \left( \frac{\sigma}{r_{pq}} \right)^{12} - \left( \frac{\sigma}{r_{pq}} \right)^{6} \right\rbrack + \epsilon \right\} H(r_{c}^{\text{WCA}} - r_{pq}). \label{eq:wca}
\end{equation}
For argon atoms, \(\sigma = \unit[0.3405]{nm}\) and
\(\epsilon = \unit[1.654 \times 10^{-21}]{J}\) are set in both potentials.

\subsection{Computation}\label{c.-computation}

Equations of motion for argon atoms are solved by the velocity Verlet
algorithm with integration time \(\Delta t = \unit[1.08 \times 10^{- 14}]{s}\)
for crystalline nanoparticles and \(\Delta t = \unit[4.3 \times 10^{- 15}]{s}\)
for amorphous nanoparticles.

All the nanoparticles prepared are initially relaxed over sufficient time steps in
the canonical (constant-temperature) ensemble at temperature \(T_\text{eq} = \unit[0.02]{K}\). After the relaxation, the
nanoparticles are brought into head-on collision at a relative impact velocity
\(v_{\text{imp}} =  v_\xi - v_\eta\) in the microcanonical (constant-energy) ensemble. The $v_\xi$ and $v_\eta$ denote the center-of-mass velocities in $z$-direction for the nanoparticles $\xi$ and $\eta$, respectively.

Our MD simulations presented hereafter are carried out by
\textsc{lammps} \cite{plimpton1995}.

\subsection{Calculation of impact
force}\label{d.-calculation-of-contact-force}

An impact force \(\mathbf{F}_{\xi\eta}\) acting on the mutual contact
surfaces formed between two nanoparticles \(\xi\) and \(\eta\) in
contact is computed by summing the individual interatomic forces
\(\mathbf{f}_{pq}\) for a pair of atoms \(p\) and \(q\) that are
positioned in separate nanoparticles \(\xi\) and \(\eta\), respectively.
The expression for the force \(\mathbf{F}_{\xi\eta}\) is given by
\begin{equation}
	\mathbf{F}_{\xi\eta} = \sum_{p \in \xi}^{}{\sum_{q \in \eta}^{}\mathbf{f}_{pq}}.
\end{equation}

The impact force \(\mathbf{f}_{pq}\) determined by the WCA
potential in Eq. (6) leads to a purely compressive impact force
\(\mathbf{F}_{\xi\eta}\) that causes deformation of the nanoparticles
during a head-on collision.

The normal component \(F_{N}\) of the impact force
\(\mathbf{F}_{\xi\eta} \) is obtained in such a way that
\(F_{N} = \mathbf{F}_{\xi\eta} \cdot \mathbf{d}_{\xi\eta}/|\mathbf{d}_{\xi\eta}|\),
where \(\mathbf{d}_{\xi\eta}\) is the instantaneous center-of-mass
distance of the colliding nanoparticles \(\xi\) and \(\eta\) in a
direction parallel to a line segment between the centers of the
colliding nanoparticles. Although the direction of
\(\mathbf{d}_{\xi\eta}\) is initially aligned with the \(z\) axis,
thermal vibrations, slip, and rotation that break the reflectional
symmetry of the system frequently result in a small deflection of the
contacting nanoparticles. The direction of
\(\mathbf{d}_{\xi\eta}\) during the collision does not necessarily
match the \(z\) axis, accordingly. The normal force \(F_{N}\) computed
by the above-stated definition is used to describe the nanoparticle
interactions.

\section{Simulation Results}\label{simulation-results}

We present simulation results in this section for impact phenomena found in our
repulsive nanoparticles. Impact forces for amorphous and monocrystalline
nanoparticles in facet and edge contacts are shown in Sec.~\ref{impact-force}. Additionally,
dynamic behaviors of deformation in the facet contact case are compared
in Sec.~\ref{dynamic-behaviors-for-facet-contact} with theoretical predictions for corresponding Hertzian spheres. For visualization of nanoparticles \textsc{vmd} \cite{humphrey1996} is utilized.

\subsection{Impact force}\label{impact-force}
\begin{figure*}[tb!]
	\begin{tabular}{cc}
		\subfloat{
		\begin{overpic}[tics=5,width=0.48\linewidth]{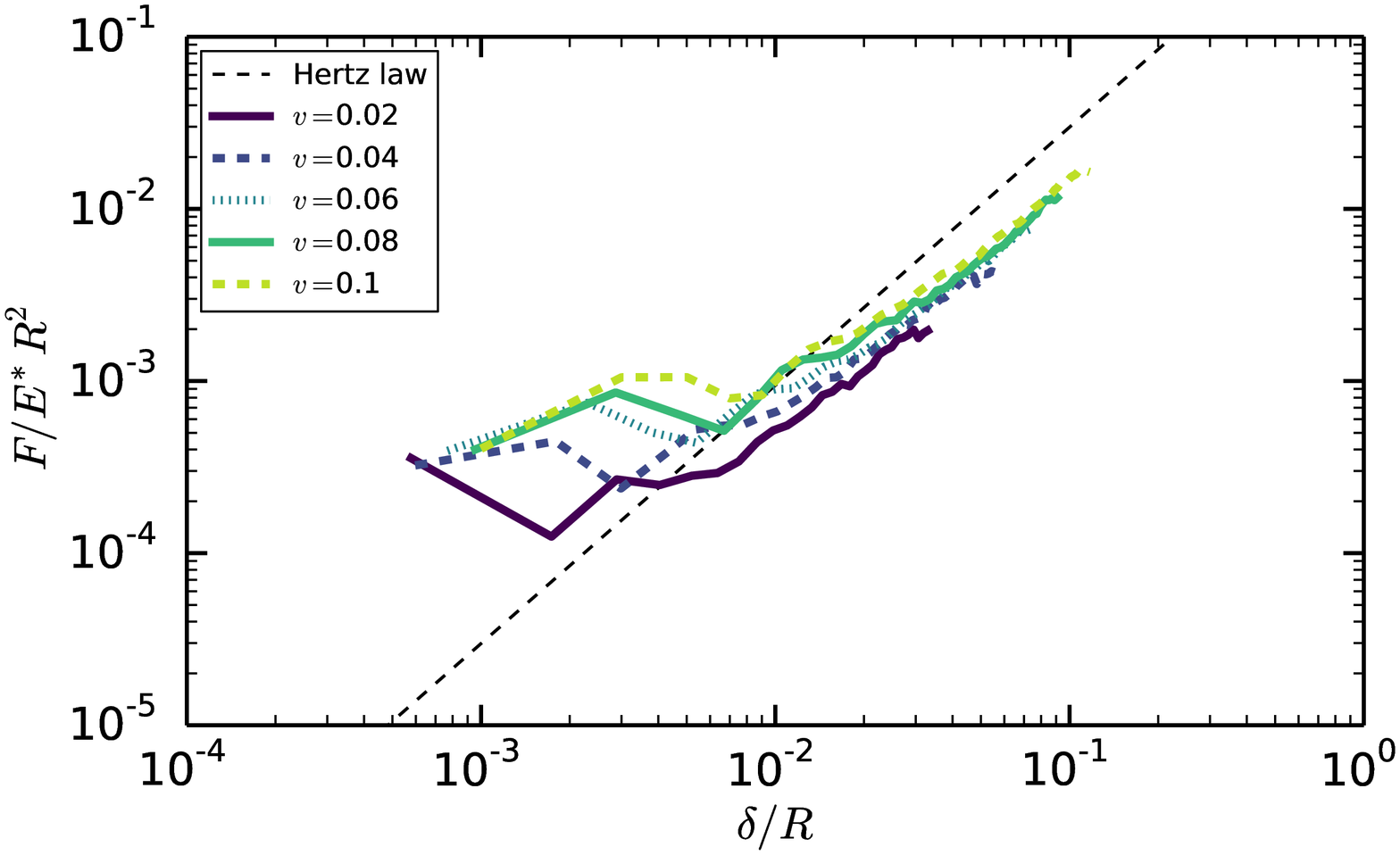}
			\put(35,51){\small (a) $R=2.3\,\mbox{nm}$}
			\put(72,35){$\eta$}
			\put(72,20){$\xi$}
			\put(67,23){\large \rotatebox{90}{$\longrightarrow$}}
			\put(67.7,34){$z$}
			\put(-1,25){\crule[white]{0.6cm}{1.5cm}}
			\put(-3,27){\rotatebox{90}{\colorbox{white}{$\Ftil$}}}
			\put(50,-2){\crule[white]{1.5cm}{0.6cm}}
			\put(51,-1){\rotatebox{0}{\colorbox{white}{ $\dtil$ }}}
			\put(21.5,38.5){\crule[white]{0.65cm}{1.15cm}}
			\put(21.5,39){\scriptsize \unit[52]{m/s}}
			\put(21.5,42){\scriptsize \unit[41]{m/s}}
			\put(21.5,45){\scriptsize \unit[31]{m/s}}
			\put(21.5,48){\scriptsize \unit[21]{m/s}}
			\put(21.5,51){\scriptsize \unit[10]{m/s}}
			\put(37,18){\small \rotatebox{-30}{$\longleftarrow$}}
			\put(44,14){\small Hertz law}
			\put(75,12){\begin{overpic}[width=0.15\linewidth,angle=90]
			{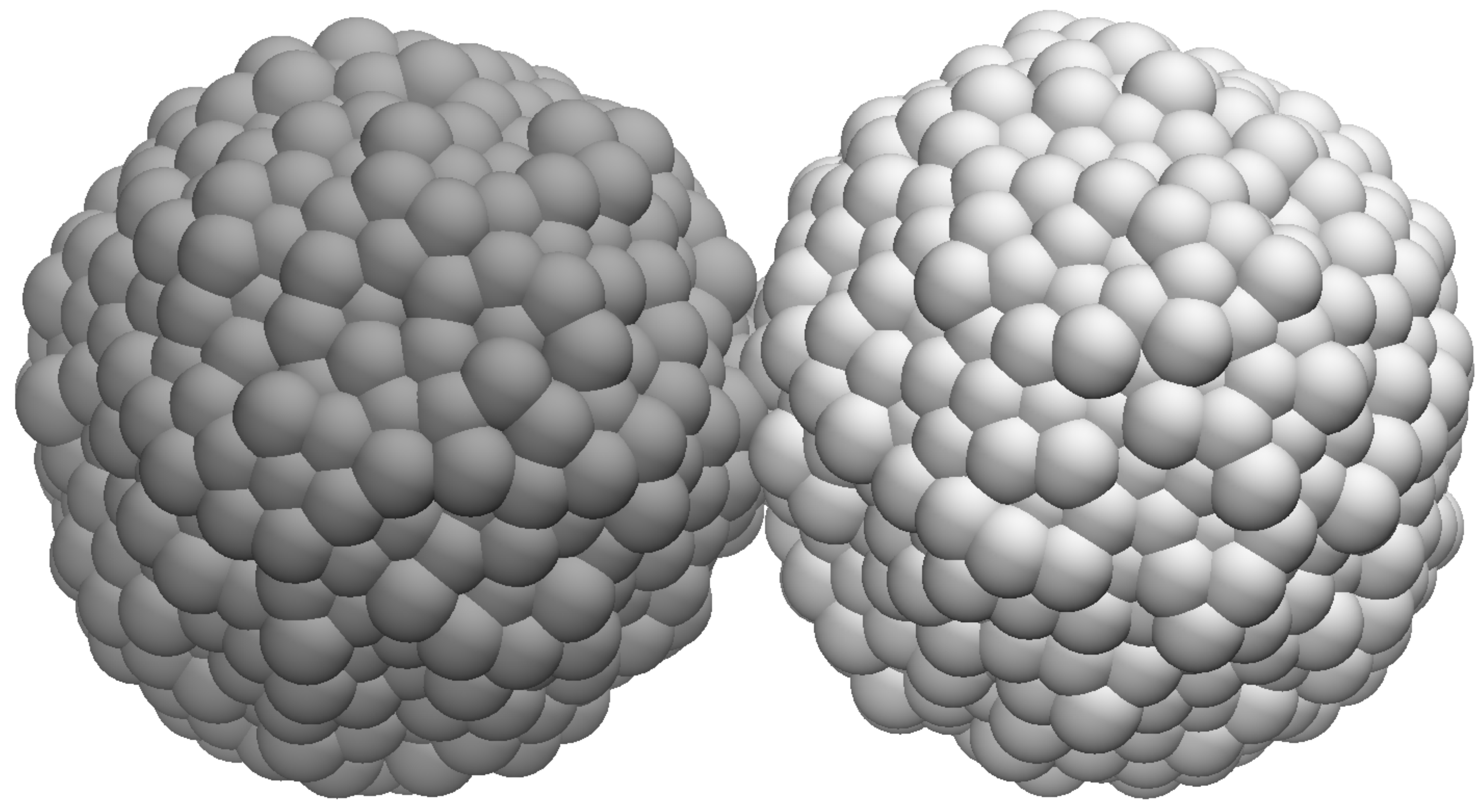}\end{overpic}}
			\label{fig:F_amorphous:2093}
		\end{overpic}}
		& \subfloat{\begin{overpic}[width=0.48\linewidth]{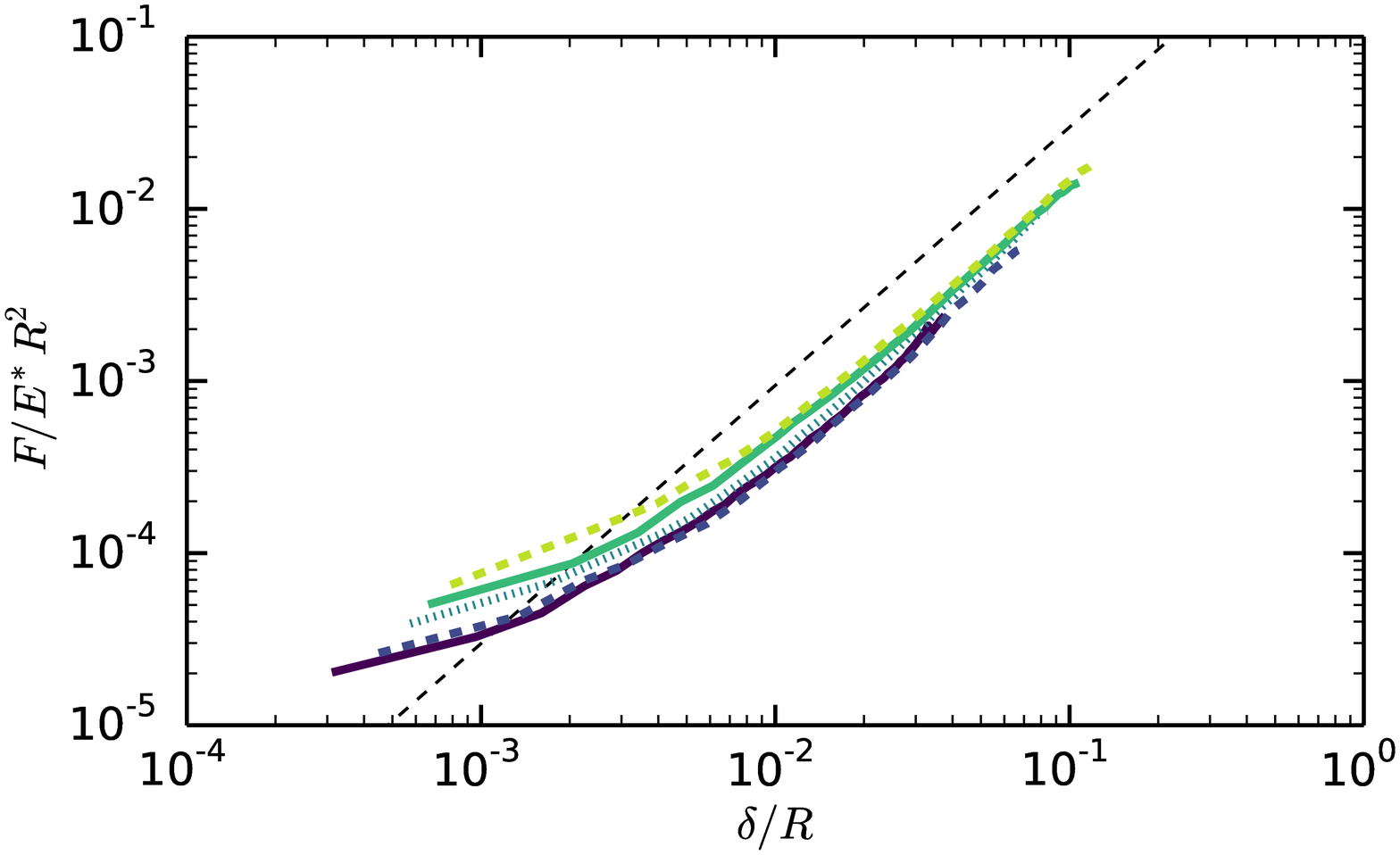}
		\put(17,51){\small (b) $R=6.2\,\mbox{nm}$}
		\put(-1,25){\crule[white]{0.6cm}{1.5cm}}
		\put(-3,27){\rotatebox{90}{\colorbox{white}{ $\Ftil$ }}}
		\put(50,-2){\crule[white]{1.5cm}{0.6cm}}
		\put(51,-1){ \rotatebox{0}{\colorbox{white}{ $\dtil$ }}}
		\put(73,12){ \begin{overpic}[width=0.15\linewidth,angle=90]
		{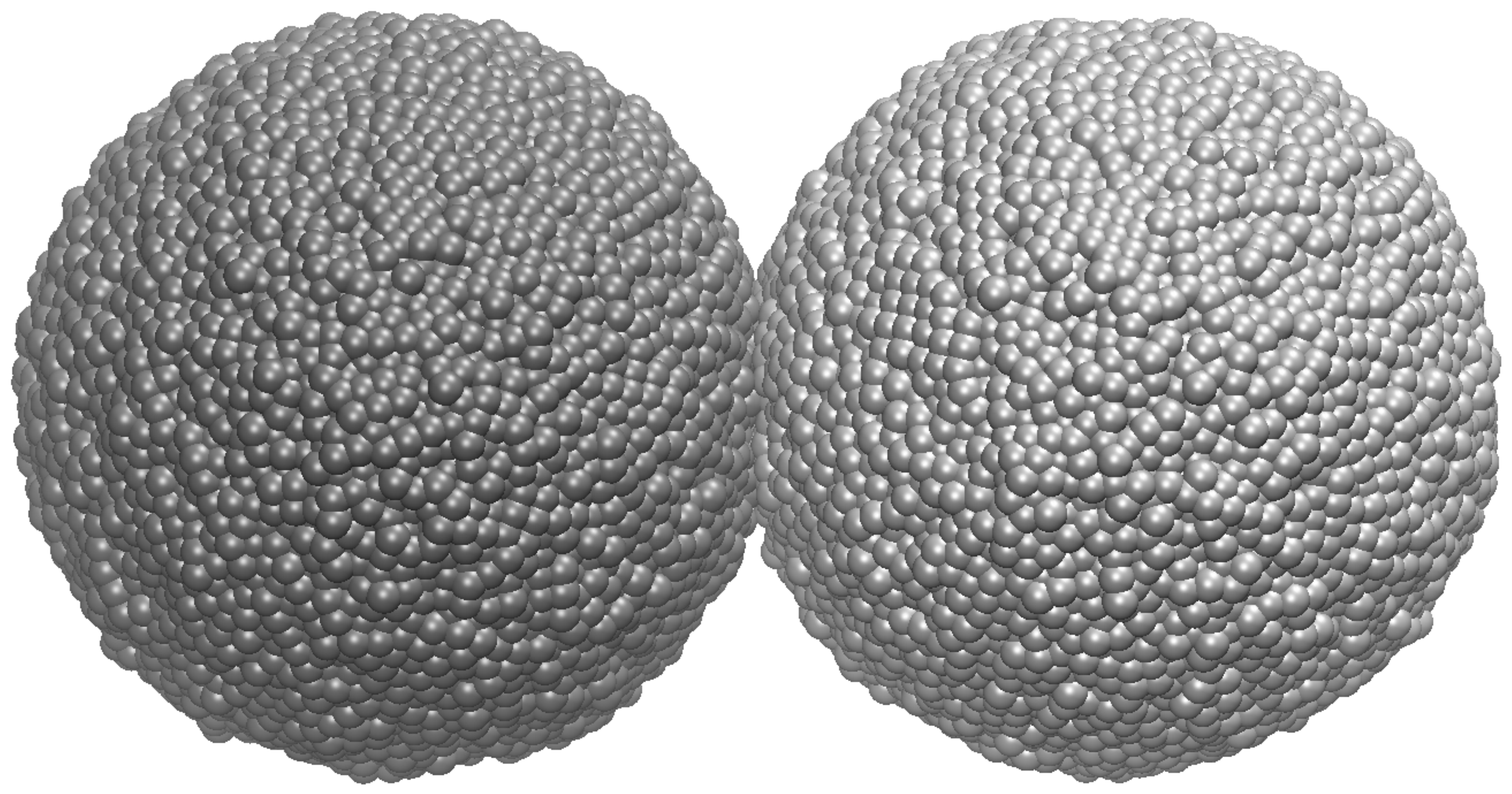}\end{overpic}}
	\label{fig:F_amorphous:44403}
	\end{overpic}}
\end{tabular}
\begin{tabular}{cc}
	\subfloat{\begin{overpic}[width=0.48\linewidth]{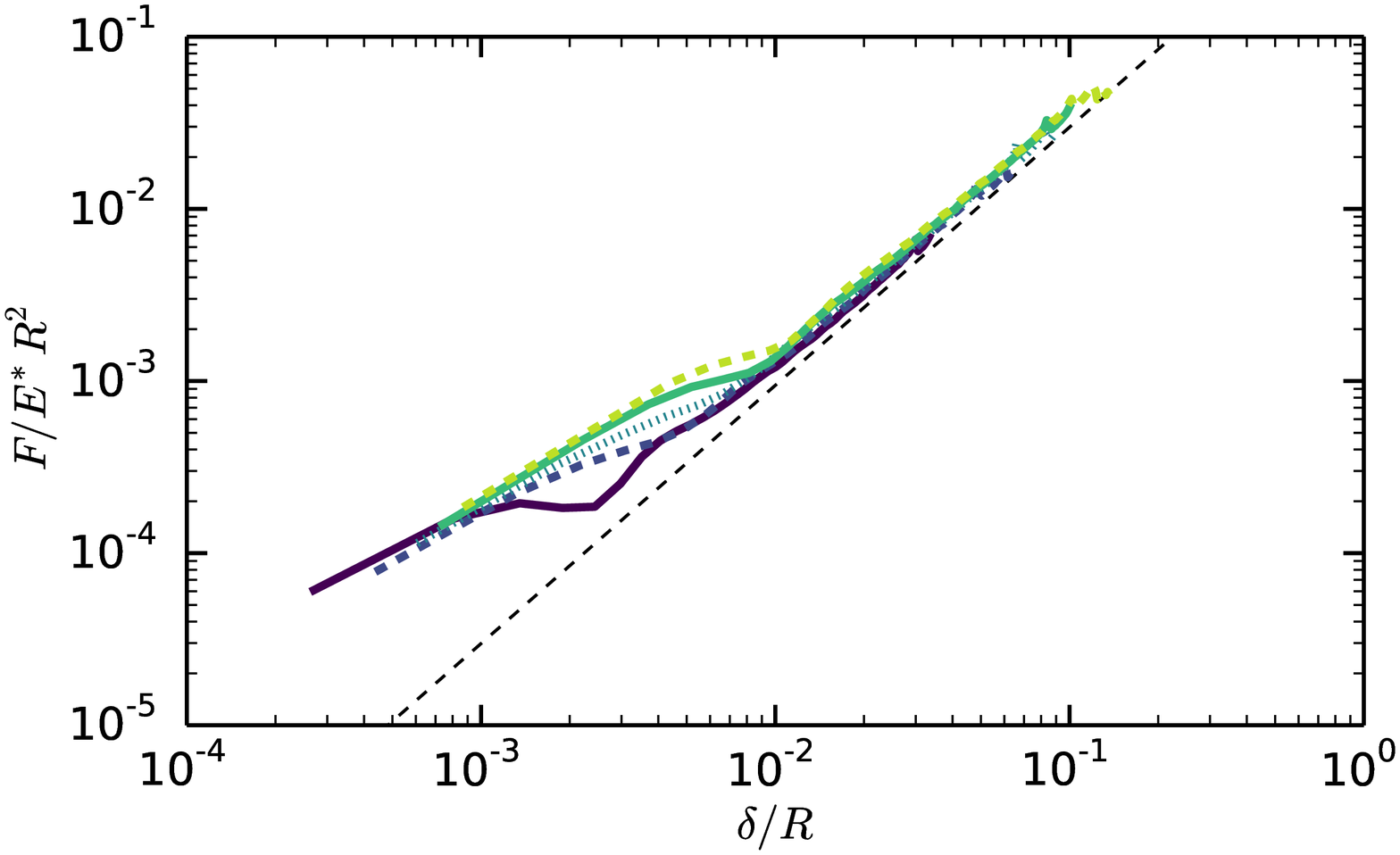}
	\put(17,51){\small (c) $R=2.7\,\mbox{nm}$}
	\put(-2,25){\crule[white]{0.6cm}{1.5cm}}
	\put(-4,27){ \rotatebox{90}{\colorbox{white}{ $\Ftil$ }}}
	\put(50,-2){\crule[white]{1.5cm}{0.6cm}}
	\put(51,-1){ \rotatebox{0}{\colorbox{white}{ $\dtil$ }}}
	\put(73,12){ \begin{overpic}[width=0.16\linewidth,angle=90]
	{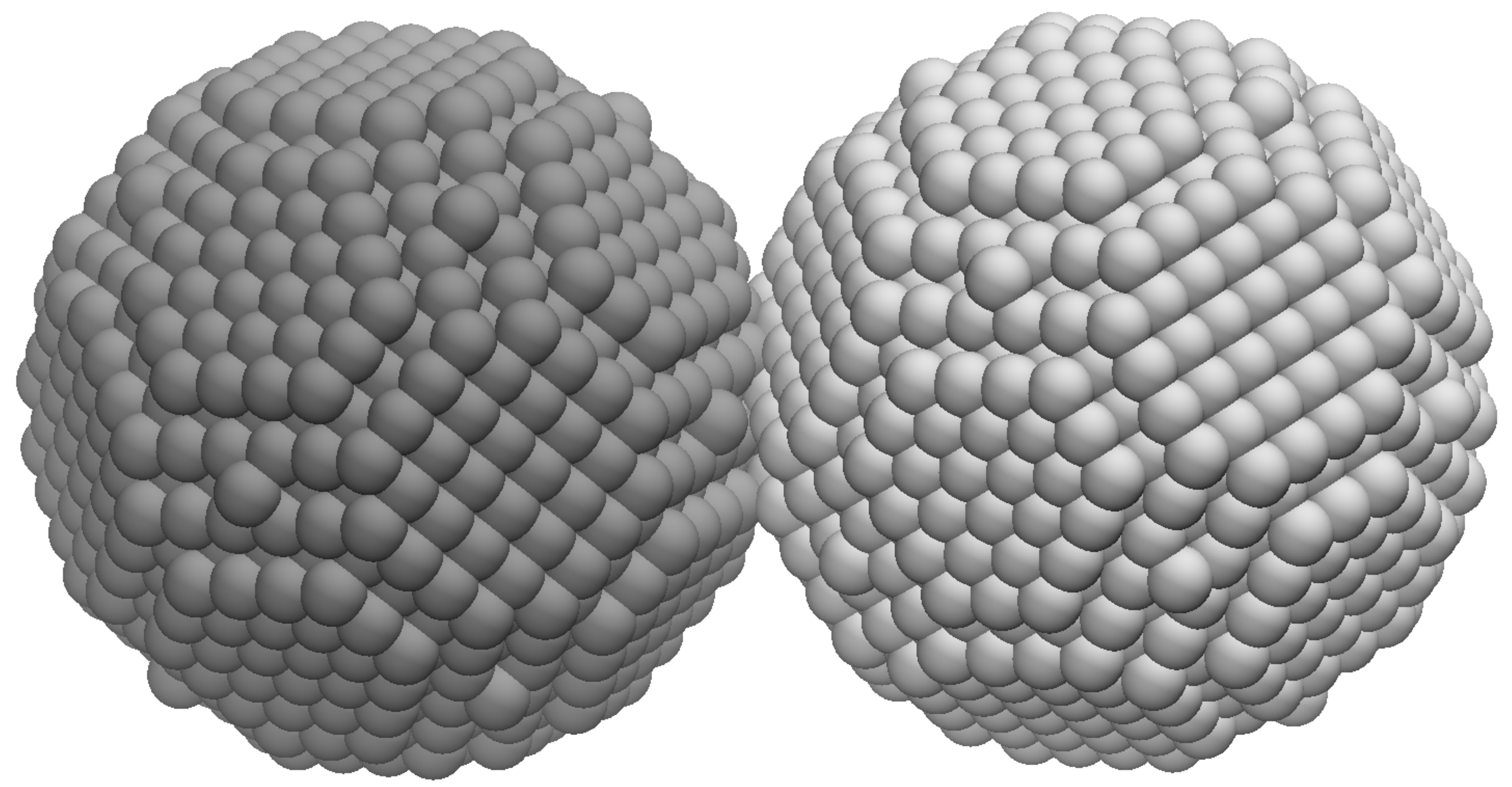}\end{overpic}}
	\label{fig:F_edge:2093}
\end{overpic}}
& \subfloat{\begin{overpic}[width=0.48\linewidth]{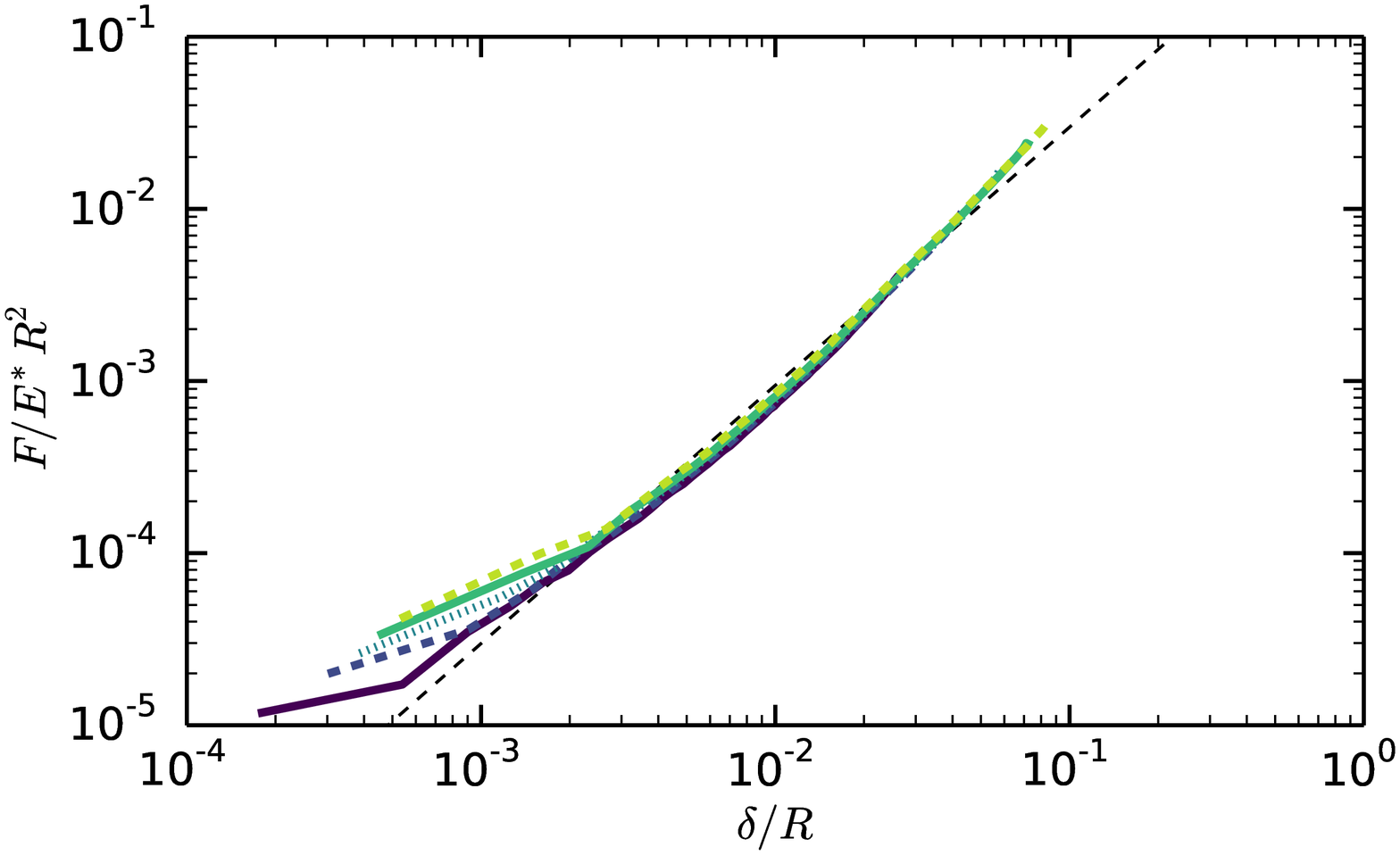}
\put(17,51){\small (d) $R=7.4\,\mbox{nm}$}
\put(-2,25){\crule[white]{0.6cm}{1.5cm}}
\put(-4,27){ \rotatebox{90}{\colorbox{white}{ $\Ftil$ }}}
\put(50,-2){\crule[white]{1.5cm}{0.6cm}}
\put(51,-1){ \rotatebox{0}{\colorbox{white}{ $\dtil$ }}}
\put(73,12){ \begin{overpic}[width=0.16\linewidth,angle=90]
{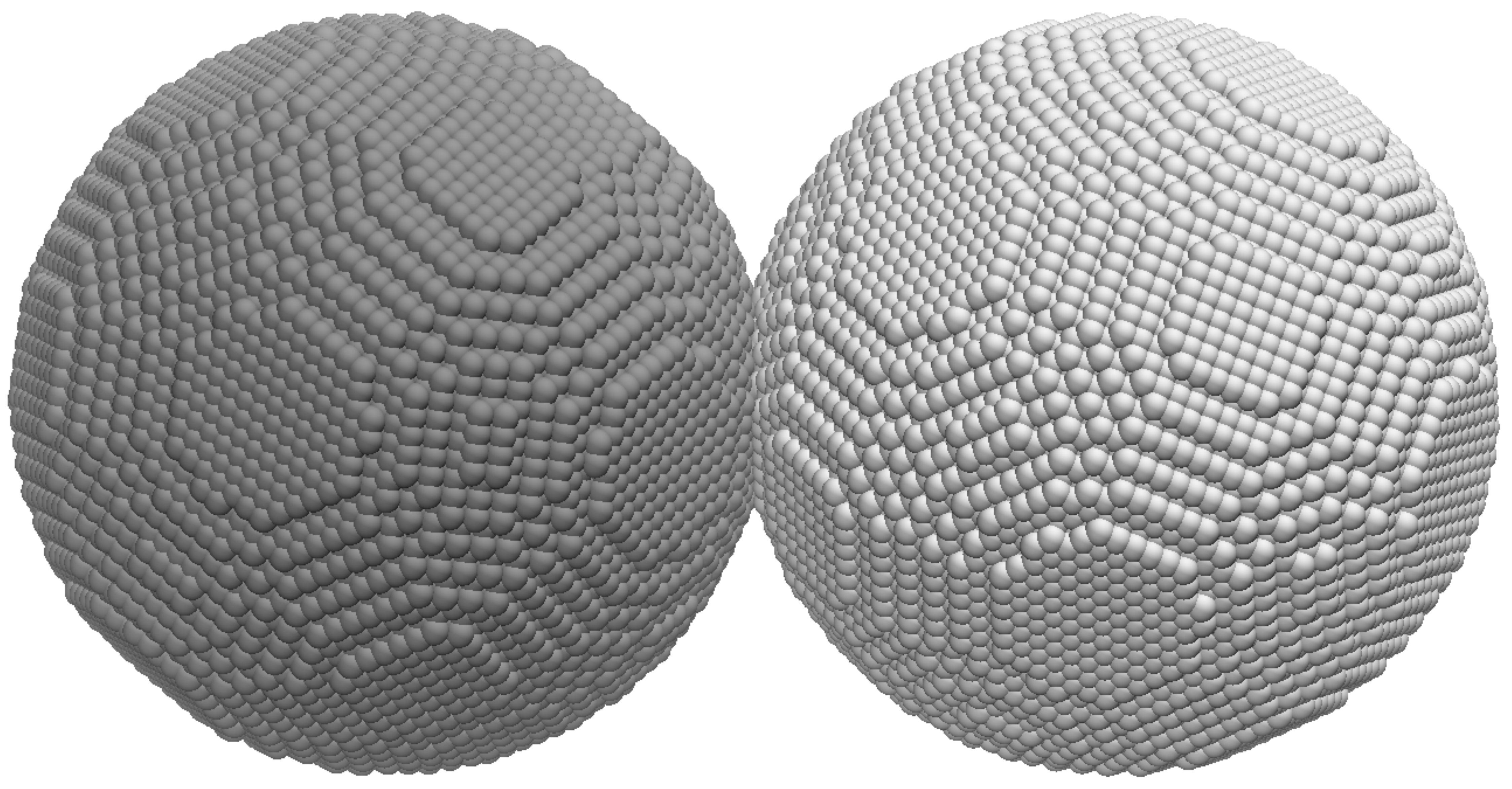}\end{overpic}}
\label{fig:F_edge:44403}
\end{overpic}}
\end{tabular}
\begin{tabular}{cc}
	\subfloat{
	\begin{overpic}[width=0.48\linewidth]{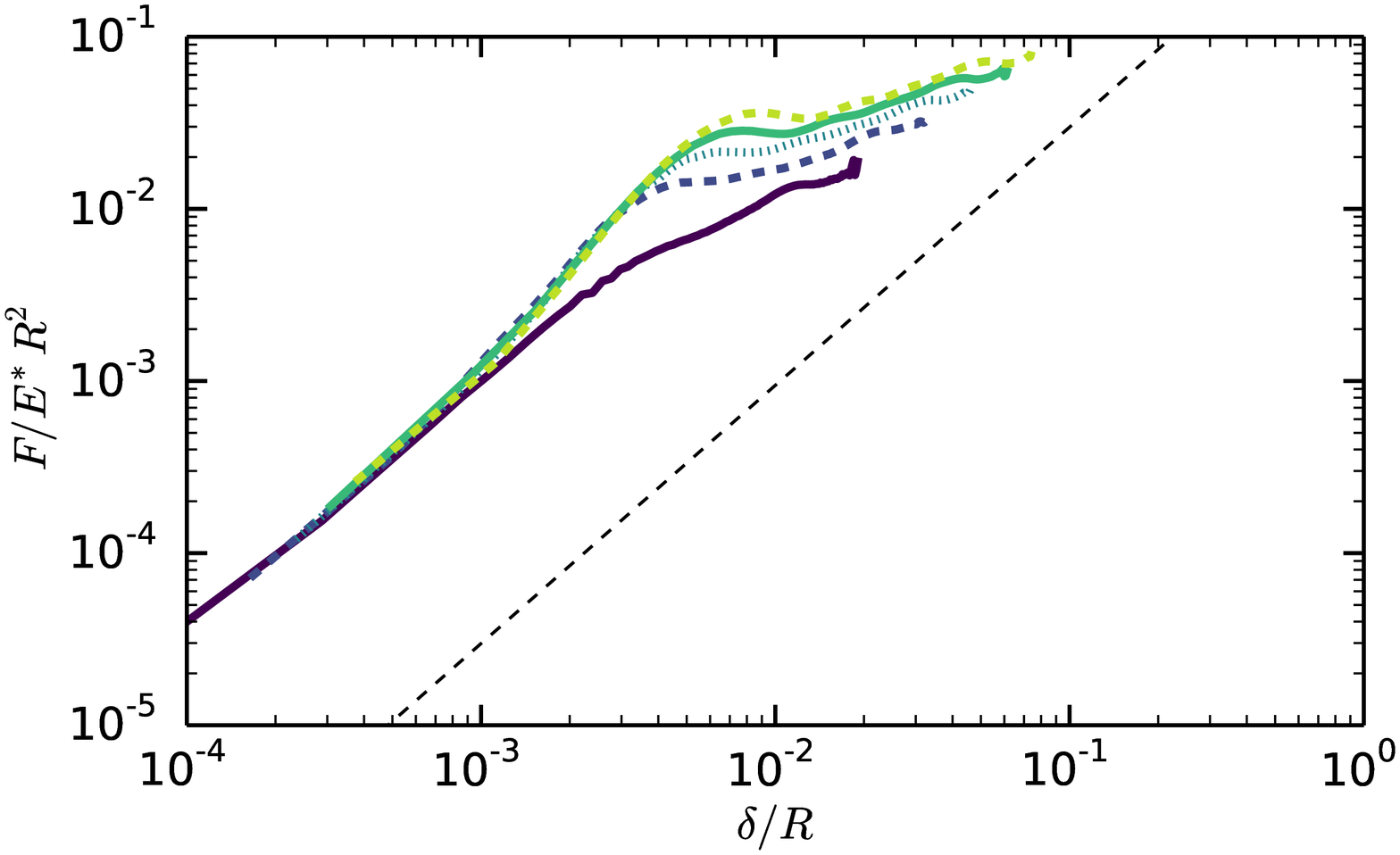}
		\put(17,51){\small (e) $R=2.5\,\mbox{nm}$}
		\put(-1,25){\crule[white]{0.6cm}{1.5cm}}
		\put(-4,27){ \rotatebox{90}{\colorbox{white}{ $\Ftil$ }}}
		\put(50,-2){\crule[white]{1.5cm}{0.6cm}}
		\put(51,-1){ \rotatebox{0}{\colorbox{white}{ $\dtil$ }}}
		\put(73,12){ \begin{overpic}[width=0.15\linewidth,angle=90]
		{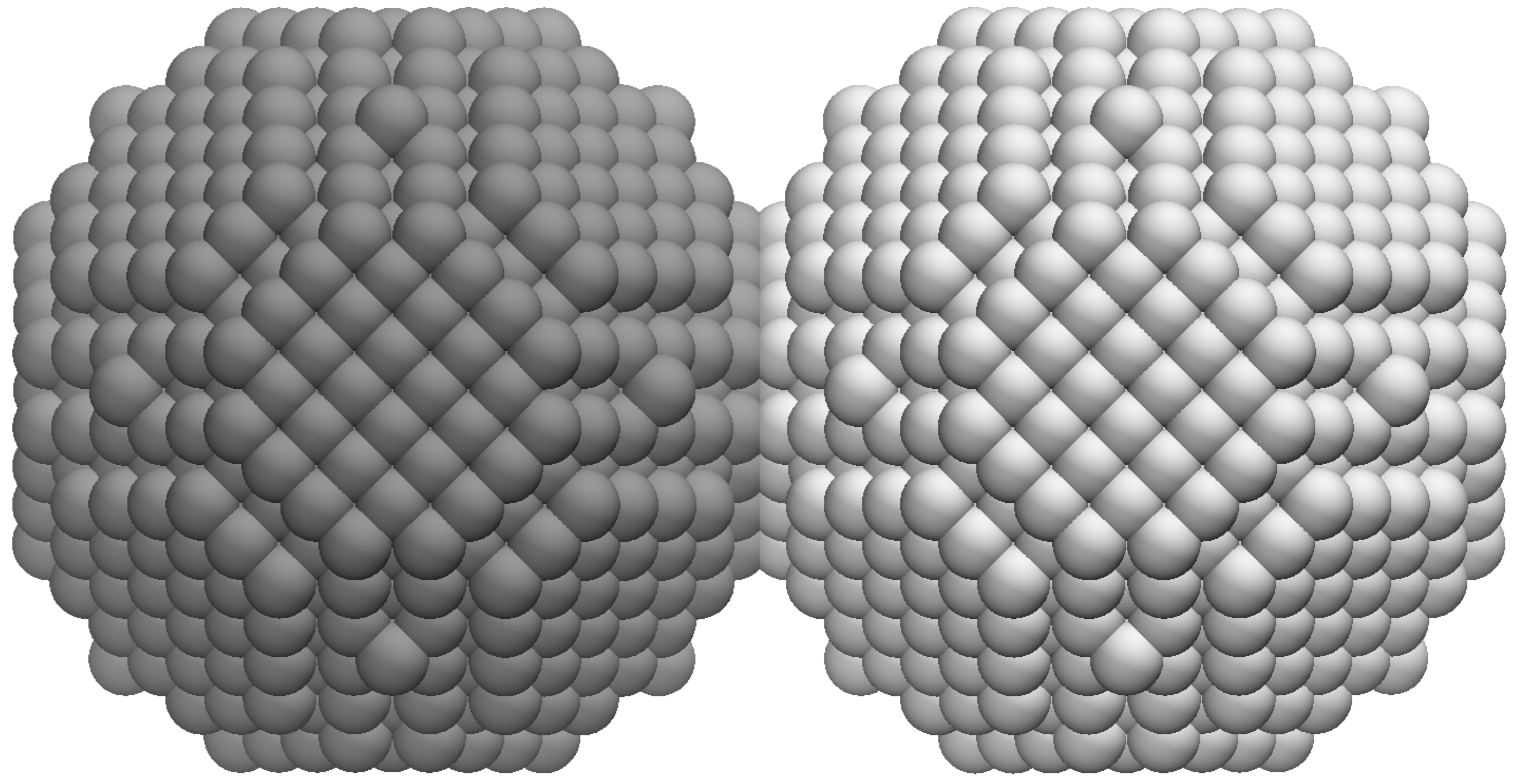}\end{overpic}}
		\label{fig:F_facet:2093}
	\end{overpic}}
	& \subfloat{\begin{overpic}[width=0.48\linewidth]{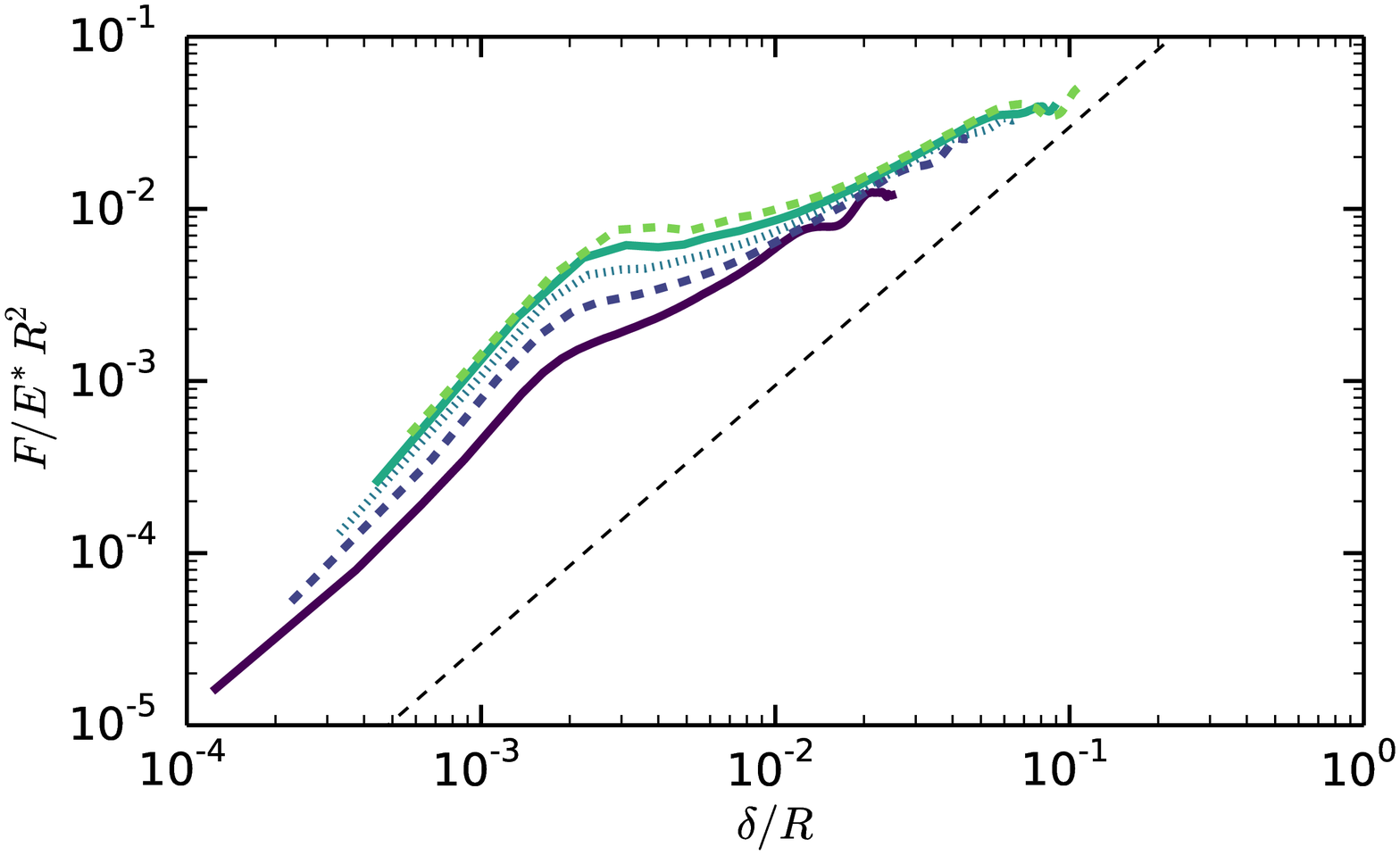}
	\put(17,51){\small (f) $R=7.3\,\mbox{nm}$}
	\put(-1,25){\crule[white]{0.6cm}{1.5cm}}
	\put(-4,27){ \rotatebox{90}{\colorbox{white}{ $\Ftil$ }}}
	\put(50,-2){\crule[white]{1.5cm}{0.6cm}}
	\put(51.5,-1){ \rotatebox{0}{\colorbox{white}{ $\dtil$ }}}
	\put(73,12){ \begin{overpic}[width=0.16\linewidth,angle=90]
	{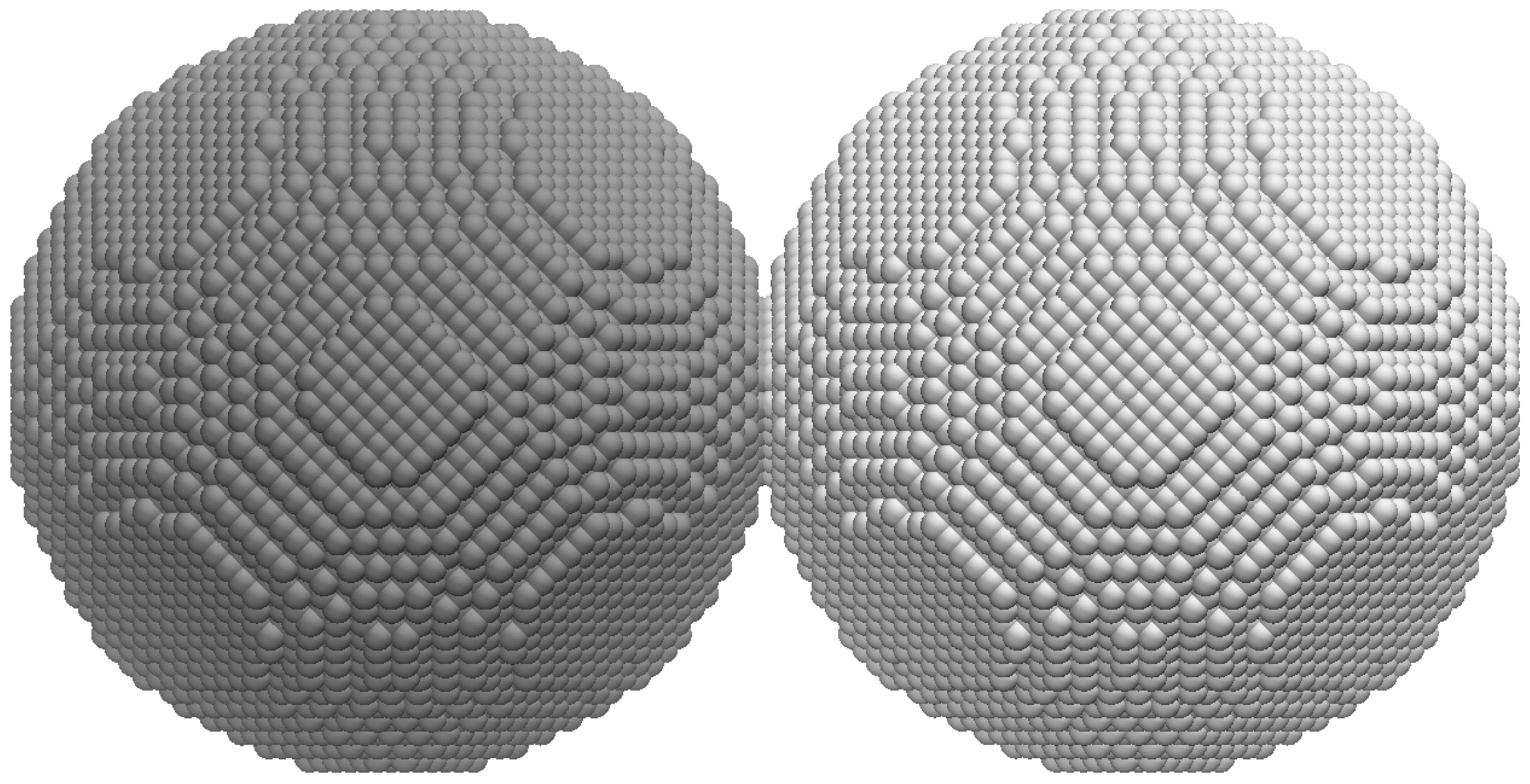}\end{overpic}}
	\label{fig:F_facet:44403}
\end{overpic}}
\end{tabular}
\caption{(Color online) Impact force \(\widetilde{F}_{N}\) for contacting
nanoparticles of radius $R$ during the loading stage, plotted as a
function of overlap \(\widetilde{\delta}\). The force depicted at each
impact velocity $\vcol$ is a set of piecewise averages over
the initial conditions. The dashed line represents the Hertzian force
for an fcc crystalline nanoparticle \cite{takato2014}. (a--b) Amorphous nanoparticles in
contact. (c--d) Crystalline nanoparticles in edge contact. (e--f)
Crystalline nanoparticles in facet contact. The snapshots of
nanoparticles shown here are taken at an impact velocity
$\vcol$ = \unit[21]{m/s}.}
\label{fig:force}
\end{figure*}

Fig.~\ref{fig:force} shows normal contact forces for amorphous nanoparticles in (a)
and (b), for fcc nanoparticles in edge contact in (c) and (d) and for
fcc nanoparticles in facet contact in (e) and (f). Each curve represents
a dimensionless normal contact force $\widetilde{F}_{N} \coloneqq F_{N}/E^{*}R^{2}$ between contacting
nanoparticles at an impact velocity \(v_{\text{imp}}\) plotted as a
function of a dimensionless overlap \(\widetilde{\delta} \coloneqq \delta/R\).
The dimensionless normal force and the dimensionless overlap are
henceforth termed a force and an overlap, respectively, for the sake of
simplicity. Two different radii \(R\) are chosen to see the size
dependence of the force. Furthermore, four different impact velocities
\(v_{\text{imp}} =\) 10, 21, 31, and \(\unit[52]{m/s}\) are selected to analyze a
dynamic effect on the force. To see the similarity between the
nanoparticles and corresponding elastic perfect spheres, the Hertzian
contact force described by Eq.~\eqref{eq:fh} is depicted as a dashed line in each plot.

The force is exhibited only in the loading stage of the collision, which
is determined by a point where the overlap \(\widetilde{\delta}\) takes
its maximum value. Although nanoparticles are highly elastic, unlike
elastic bodies in contact, the force in the unloading stage mostly does
not follow back the path of the force-overlap curve taken in the loading
stage. This is because of irreversible processes. A part of the initial kinetic energy is almost always converted to thermal energy during the collision.

All plots in~Fig.~\ref{fig:force} have two compression regimes where the rates of
change differ noticeably in impact force. In a curve for \(R = \unit[2.5]{nm}\)
and \(v_{\text{imp}} = \unit[52]{m/s}\) in Fig.~\ref{fig:force}(e), for instance, its critical
overlap \({\widetilde{\delta}}_{c}\) is identified as a kink observed at
around \(\widetilde{\delta} \sim 10^{-2}\), which varies depending on
the contact type, size, and impact speed. The force below the critical
overlap increases at some constant rate as the overlap increases, and
the force above the critical overlap increases further at another rate.

In the high compression regime,
\(\widetilde{\delta} \gtrsim \dtil_{c}\), the edge contact in Fig.~\ref{fig:force} (c)
and (d) shows good agreement with the Hertz contact theory, and the
amorphous nanoparticles in Fig.~\ref{fig:force} (a) and (b) recover the slope of the
Hertz contact force. The facet contact, however, turns out to be
distinct from the Hertz contact force in both slope and magnitude.

In the low compression regime,
\(\widetilde{\delta} \lesssim \dtil_{c}\), all the forces appear to increase at a rate similar to the Hertz contact force but show large
departures in magnitude. The edge contact and amorphous
nanoparticles in the same figures possess a gentler mechanical response
to impact loads, but the facet contact shows a rapid increase in force
in (e) and (f).

Impact forces for the aforementioned surface roughnesses and compression levels are summarized in Table \ref{table:force}, and we will show more details in the contact types separately.

\begin{table}[tb!]
	\centering
	\caption{Impact force for contact surface types and overlap ranges.}
	\label{table:force}
\begin{tabular}{c|ccc}
	& \multirow{2}{*}{Amorphous nanoparticle}  & \multicolumn{2}{c}{Crystalline nanoparticle} \\
	 & &  Edge contact & Facet contact \\  \hline \hline
	Large compression $\dtil > \dtil_c$ & Hertz & Hertz & Non-Hertz \\
	Small compression $\dtil < \dtil_c$ & Non-Hertz & Non-Hertz & Non-Hertz
\end{tabular}
\end{table}

\subsubsection{Impact between amorphous
nanoparticles}\label{i.-impact-between-amorphous-nanoparticles}

The exteriors of the amorphous nanoparticles in Figs.~\ref{fig:force}(a) and (b) look more
spherical than those of the crystalline nanoparticles in Figs.~1(a) and
(b). Although individual atoms randomly stacked on the surface form a
small asperity, the arrays of atomic steps and associated sharp edges
seen in the crystalline nanoparticles are absent due to the disordered
arrangement. The facet-free nanoparticle is therefore expected to
recover a Hertzian-like force curve with the \(3/2\) power of overlap.

Impact forces \({\widetilde{F}}_{N}\) of the large amorphous
nanoparticles in Fig.~\ref{fig:force} (b) at high compression
\(\widetilde{\delta} \gtrsim \dtil_{c}\) resemble the Hertz contact
force in the sense that \({\widetilde{F}}_{N}\) grows with overlap nonlinearly at a
rate \(\sim 1.5\) though there is a dynamic effect, that is, the forces
increase at a certain overlap fixed as impact velocity increases. The
nanoparticles in the inset of Fig.~\ref{fig:force}(b)~have a flat contact surface
developed during the collision. The expanding contact surface area seems
consistent with the result in Ref.~\cite{kim2010}, which reported an expanding contact surface area of colliding polymer nanoparticles having a relatively smooth
surface like our amorphous nanoparticles. This area is in agreement with that of corresponding Hertz spheres in
Eq.~(2) when the compression is high enough. This contact surface
expansion induced by compression as expressed in Eq.~(2) is critical to recover the
Hertzian contact force.

The smaller amorphous nanoparticles in Fig.~\ref{fig:force}(a) show an impact
response similar to that of the large ones, but some noise is present.
The contact surface of the small nanoparticles in compression involves
only a small number of atoms due to their size, \(R = \unit[2.3]{nm}\), the
smallest among our nanoparticles.

\subsubsection{Impact on sharp crystal
edges}\label{ii.-impact-on-sharp-crystal-edges}

We consider cases where two monocrystalline nanoparticles collide such that they first come into
contact on at least one sharp edge on either of the nanoparticles.
Fig.~\ref{fig:force}(c) and (d) illustrate the collisions for fully compressed
nanoparticles that are randomly oriented before collision. As noticed,
the edge collision case obviously has a contact area smaller than the
\{001\} facet displayed in Fig.~\ref{fig:force}(e). This allows the contact surface
area in edge contact to expand as the overlap increases.

This contact force is in excellent agreement with the Hertz contact
force, as shown in Figs.~\ref{fig:force}(c) and (d). The impact force is well aligned
with the Hertzian force when both the nanoparticles are compressed
sufficiently, \(\widetilde{\delta} > {\widetilde{\delta}}_{c}\). In this
regime, the magnitude of the impact force has almost negligible velocity
dependence. Regardless of the impact velocity variation between 10 and
\(\unit[52]{m/s}\), the forces for the given velocities are all the same up to
their maximum overlaps, which now depend on the impact velocity.

\subsubsection{Impact on crystal
facets}\label{iii.-impact-on-crystal-facets}

The contact surface for the facet contact in the inset of Fig.~\ref{fig:force}(e) is
obviously distinct from those in the edge contact and amorphous
nanoparticle cases. The presence of large surfaces on crystalline
nanoparticles and the impact on the facets completely alter the dynamic
response of the colliding nanoparticles. Consequently, the impact force
between the nanoparticles has a behavior unlike the Hertz contact force
in Eq.~\eqref{eq:fh}.

When compression of the nanoparticles is low
\(\widetilde{\delta} < {\widetilde{\delta}}_{c}\), all impact forces
\({\widetilde{F}}_{N}\) at various velocities in Fig.~\ref{fig:force}(e) are the same,
and are considerably higher than their corresponding Hertzian forces.
Each force then increases at a lower rate as the compression increases
further. The slope of the force-overlap curve in the large overlap
region, \(\widetilde{\delta} > {\widetilde{\delta}}_{c}\), is notably
lower than \(n = 3/2\) for the Hertz contact force, shown as a dashed
line in Figs.~\ref{fig:force}(a--e). Furthermore, the force in the same overlap range has
a relatively strong velocity dependence, and faster nanoparticles
experience stronger repulsive forces.

The larger nanoparticles in Fig.~\ref{fig:force}(f) behave qualitatively in the same
way as seen in the smaller nanoparticles, that is, the impact force has
a higher rate of change at
\(\widetilde{\delta} < {\widetilde{\delta}}_{c}\) and a lower rate at
\(\widetilde{\delta} > {\widetilde{\delta}}_{c}\). Large nanoparticles are expected to approach Hertz spheres since as nanoparticle size is increased, the
facet area progressively becomes smaller and the role of the facet would
vanish in the limiting case of \(R \rightarrow \infty\). However, these
nanoparticles of \(R = \unit[7.3]{nm}\) bear no semblance of Hertzian spheres.

\subsection{Dynamic behaviors for facet
contact}\label{dynamic-behaviors-for-facet-contact}

\subsubsection{Contact surface area}\label{i.-contact-surface-area}

The contact surface area developed for Hertzian spheres under
compression is described by Eq.~(2) and it gives rise to nonlinearity in
its mechanical response. In the case of the facet collision, the \{001\}
facets are chosen for surfaces to be contacted, and the facet contact
resulted in the non-Hertzian force, as already seen. Thus, it is
interesting to see how the mutual contact surface evolves as the
nanoparticles are compressed. The contact surface area for nanoparticles
of several sizes is examined and compared with the Hertzian sphere case.

Fig.~\ref{fig:force}(e) shows a representative snapshot for faceted nanoparticles with
their \{001\} facets in contact at maximum compression, and their
colliding surfaces possess relatively flat square-like areas. We adopt a
new measure to quantify a contact radius \(a\) for our nanoparticles in
place of a contact surface area. The contact radius \(a\) is an
effective radius for an actual contact surface area that equals a
circular surface area \(\pi a^{2}\). The actual contact surface of the
contacting nanoparticles is not a circle; however, it is convenient to
regard the square-like contact surface as a circular surface of radius
\(a\) when compared to the radius of the mutual contact surface area
formed between the Hertzian spheres. The contact radius \(a\) for
nanoparticles is computed by  \cite{vergeles1997}
\begin{equation}
\quad a^2 = \frac{1}{N_{s}^{2}}\sum_{i = 1}^{N_{s}}{\sum_{j = 1}^{N_{s}}\left\lbrack \left( x_{i} - x_{j} \right)^{2} + \left( y_{i} - y_{j} \right)^{2} \right\rbrack},
\end{equation}
where \(x_{i}\) (\(y_{i}\)) and \(x_{j}\) (\(y_{j}\)) are the \(x\)
(\(y\)) component of \(i\)-th and \(j\)-th atom coordinates on the same
contact surface, respectively.
\(N_{s}\) denotes the total number of atoms involved in contact.

\begin{figure}[tb!]
	\centering
	\begin{overpic}[width=0.6\linewidth]{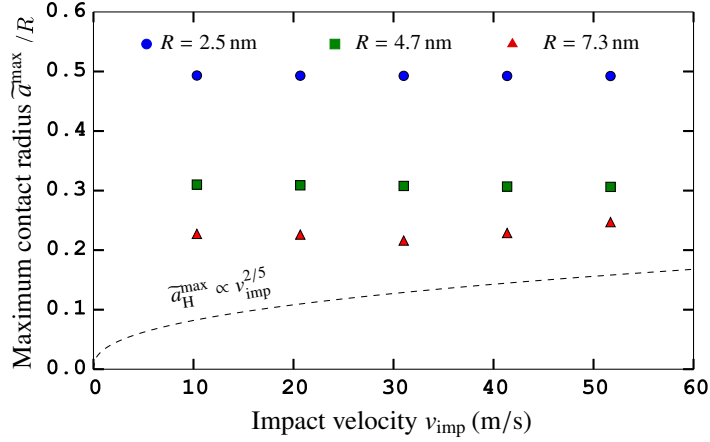}
		\put(3,8){\rotatebox{90}{\colorbox{white}{Maximum contact radius $\atilmax$}}}		\put(35,1){\colorbox{white}{Impact velocity $\unit[\vcol]{(m/s)}$}}
		\put(20,51.5){\myCirc}
		\put(46,51.5){\mySqua}
		\put(67,51.5){\myTri}
		\put(24,52){\footnotesize $R=\unit[2.5]{nm}$}
		\put(50,52){\footnotesize $R=\unit[4.7]{nm}$}
		\put(75,52){\footnotesize $R=\unit[7.3]{nm}$}
		\put(25,18.){\footnotesize \rotatebox{9}{$\atilmax_\text{H} \propto \vcol^{2/5}$}}
	\end{overpic}
	\caption{(Color online) The dimensionless maximum contact radius
	\({\tilde{a}}^{\max}\) computed from our simulation data remains
	constant over the quasi-elastic collision range. The dashed line is
	predicted by the Hertz contact theory, which is independent of size $R$.} \label{fig:contRad}
\end{figure}

The nondimensionalized contact radii \(\tilde{a} \coloneqq a/R\) of three
different sized nanoparticles computed by Eq.~(8) at their maximum
compression are plotted against the impact velocity in Fig.~2.

The plateaus in the figure clearly indicate that the maximum contact
radii \({\widetilde{a}}^{\max}\) remain unchanged. In other words, the
facet contact does not allow a contact surface expansion no matter how
forcefully the nanoparticles collide, provided that the impact occurs
within the range of the quasi-elastic collision regime. Facet sizes are
nearly comparable to the radii of small nanoparticles. In particular, the
contact radius for the \(R = \unit[2.5]{nm}\) nanoparticle reaches \(50\%\) of
its radius. By contrast, the surface area for Hertzian spheres expands
with increasing impact velocity depicted by a dashed line  based on Eq. (2), as seen in Fig.~2.

\subsubsection{Compression}\label{ii.-compression}
Another mechanical property examined here is how much the nanoparticles
are compressed at maximal contact. Colliding elastic spheres show a
velocity dependence on the dimensionless maximum overlap
\({\widetilde{\delta}}^{\max} \propto {v_{\text{imp}}}^{\alpha}\), where
\(\alpha = 4/5\) for the Hertz, but no size dependence according to
Eq.~(3). The exponent \(\alpha = 4/5\) of the impact velocity
\(v_{\text{imp}}\) equals the ratio of the exponent, 2, of the impact
velocity in kinetic energy \((1/2)M^{*}{v_{\text{imp}}}^{2}\) to the
exponent \(n + 1 = 5/2\) of the overlap in elastic energy
\(\kappa_{H}\delta^{n + 1}\) stored in the deformed Hertzian spheres,
i.e., \(\alpha = 2/(n + 1) = 4/5\). A different value of the exponent
\(\alpha\) in the facet contact case would be anticipated since its
force does not obey the Hertzian force in Figs.~\ref{fig:force}(e) and (f) and the
power of the force law, $n$, is likely different from 3/2.

\begin{figure}[tb!]
	\centering
	\begin{overpic}[width=0.6\linewidth]{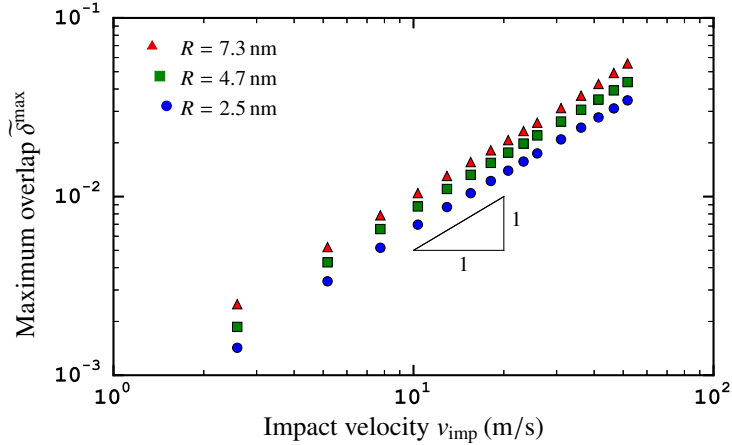}
		\put(61,23){\footnotesize 1}
		\put(68,29){\footnotesize 1}
		\put(16.3,52){\myTri}
		\put(20,48){\mySqua}
		\put(20,44){\myCirc}
		\put(24,52){\footnotesize $R=\unit[7.3]{nm}$}
		\put(24,48){\footnotesize $R=\unit[4.7]{nm}$}
		\put(24,44){\footnotesize $R=\unit[2.5]{nm}$}
		\put(29,-3){\crule[white]{5.5cm}{0.7cm}}
		\put(35,1){Impact velocity $\unit[\vcol]{(m/s)}$}
		\put(0,13){\crule[white]{0.65cm}{3.7cm}}
		\put(0,13){\rotatebox{90}{\colorbox{white}{Maximum overlap $\dtilmax$}}}
	\end{overpic}
	\caption{(Color online) The dimensionless maximum overlap
	\({\tilde{\delta}}^{\max}\) obtained from our simulations is
	proportional to $\vcol$. The Hertz contact theory predicts
	a velocity dependence in
	\({\tilde{\delta}}^{\max} \propto v_{\text{imp}}^{4/5}\).}
	\label{fig:maxComp}
\end{figure}

Fig.~\ref{fig:maxComp} presents the maximum overlap \({\widetilde{\delta}}^{\max}\) of
the nanoparticles. The obtained maximum overlap for three nanoparticle
sizes has a size dependence, and the smaller nanoparticles are less
deformable. The maximum overlap of the nanoparticles exhibits a velocity
dependence and appears to be proportional to the impact velocity. This
result differs from the \(4/5\) power of the velocity that the Hertz
contact theory yields. This deviation in the exponent demonstrates
further evidence of the non-Hertzian contact force observed in the facet
contact.

\section{Discussion}\label{discussion}

\subsection{Critical overlap and contact surface area at low
compression}\label{a.-critical-overlap-and-contact-surface-area-at-low-compression}

We defined the critical overlap \({\widetilde{\delta}}_{c}\) as an
overlap where the dynamic response of the colliding nanoparticles
undergoes a sudden change in slope of a force-overlap curve. This value
can reduce the somewhat complex force behavior into two regimes in which
the forces grow at different rates in a comparatively smooth manner. It
then raises questions about what makes such a change in force at a
particular overlap and what influences force behaviors above and below
the critical overlap.

Fig.~4(a) is a force-overlap plot for facet
contact. In addition, the number $N_s$ of atoms involved in contact is also
included in the plot in order to see how evolution of the contact
surface influences the force. The dimensionless impact force \(\Ftil\) and number \(N_{s}\)
in the plot are obtained by averaging over different initial conditions.

The number \(N_{s}\) of atoms in contact rises and then saturates as the
overlap increases. At saturation, all atoms residing on the very first
atomic layer of the facet participate in contact. The atomic layers are
stacked in the {[}001{]} direction, and the atoms in the second layer
are clearly visible from the exteriors of the nanoparticles in
Fig.~1(e). Nevertheless, the second layers never come into contact with
the other nanoparticle; hence, further growth of the contact area does
not occur.

\begin{figure}[tb!]
		\subfloat[Facet contact, $R=2.5\,\mbox{nm}$ ]{\begin{overpic}[tics=5,width=0.48\linewidth]{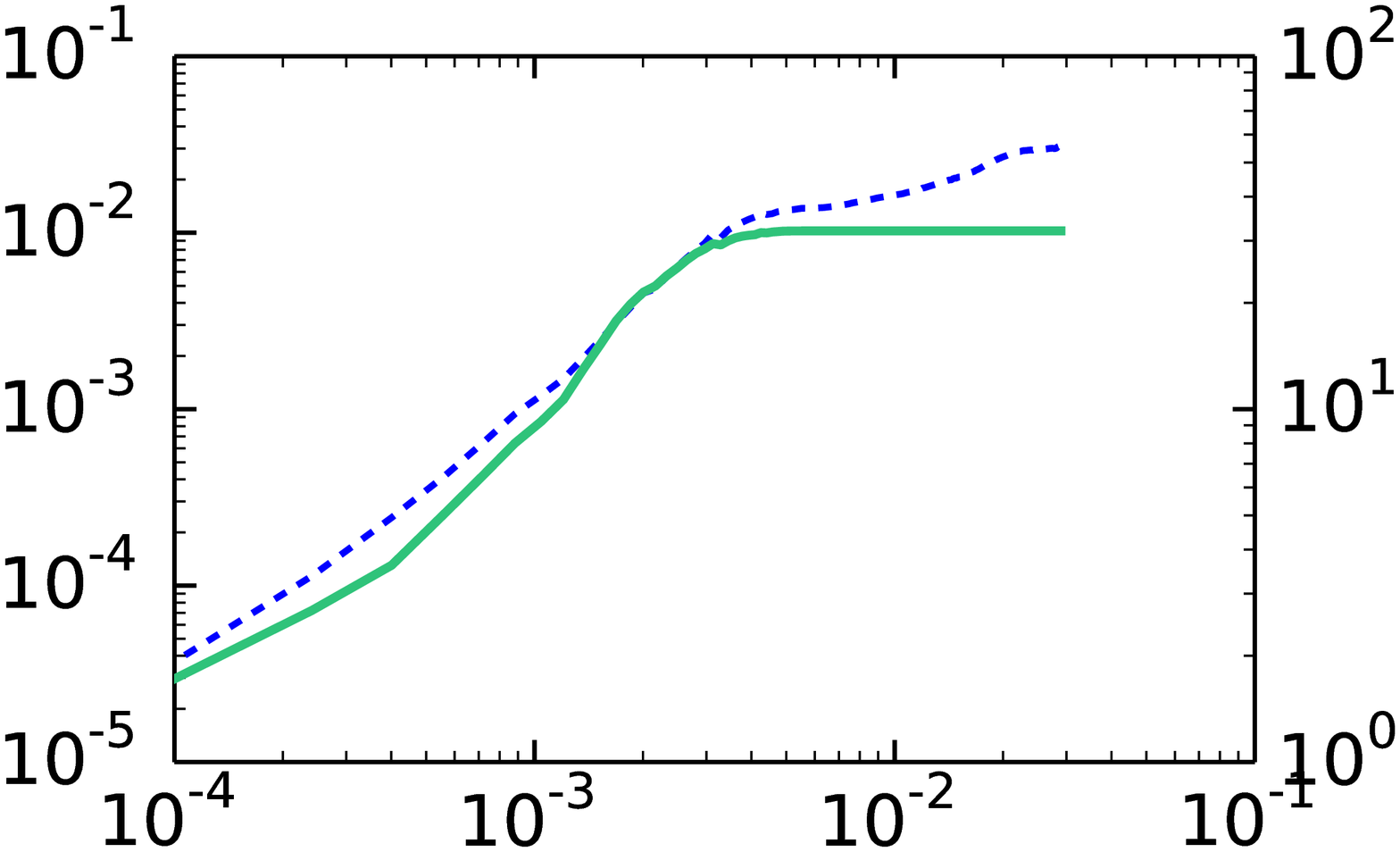}
		\put(50,32){$\dtil_c$}
		\put(-3,30){\rotatebox{90}{\colorbox{white}{\textcolor{myblue2}{$\Ftil$}}}}
		\put(47,0){\rotatebox{0}{\colorbox{white}{$\dtil$}}}
		\put(93.5,30){\rotatebox{90}{\colorbox{white}{\textcolor{mygreen2}{$N_s$}}}}
		\label{fig:FN_facet:2}
	\end{overpic}}
		\hspace{1mm}
		\subfloat[Edge contact, $R=2.7\,\mbox{nm}$ \label{fig:FN_edge:2}]{
		\begin{overpic}[tics=5,width=0.48\linewidth]{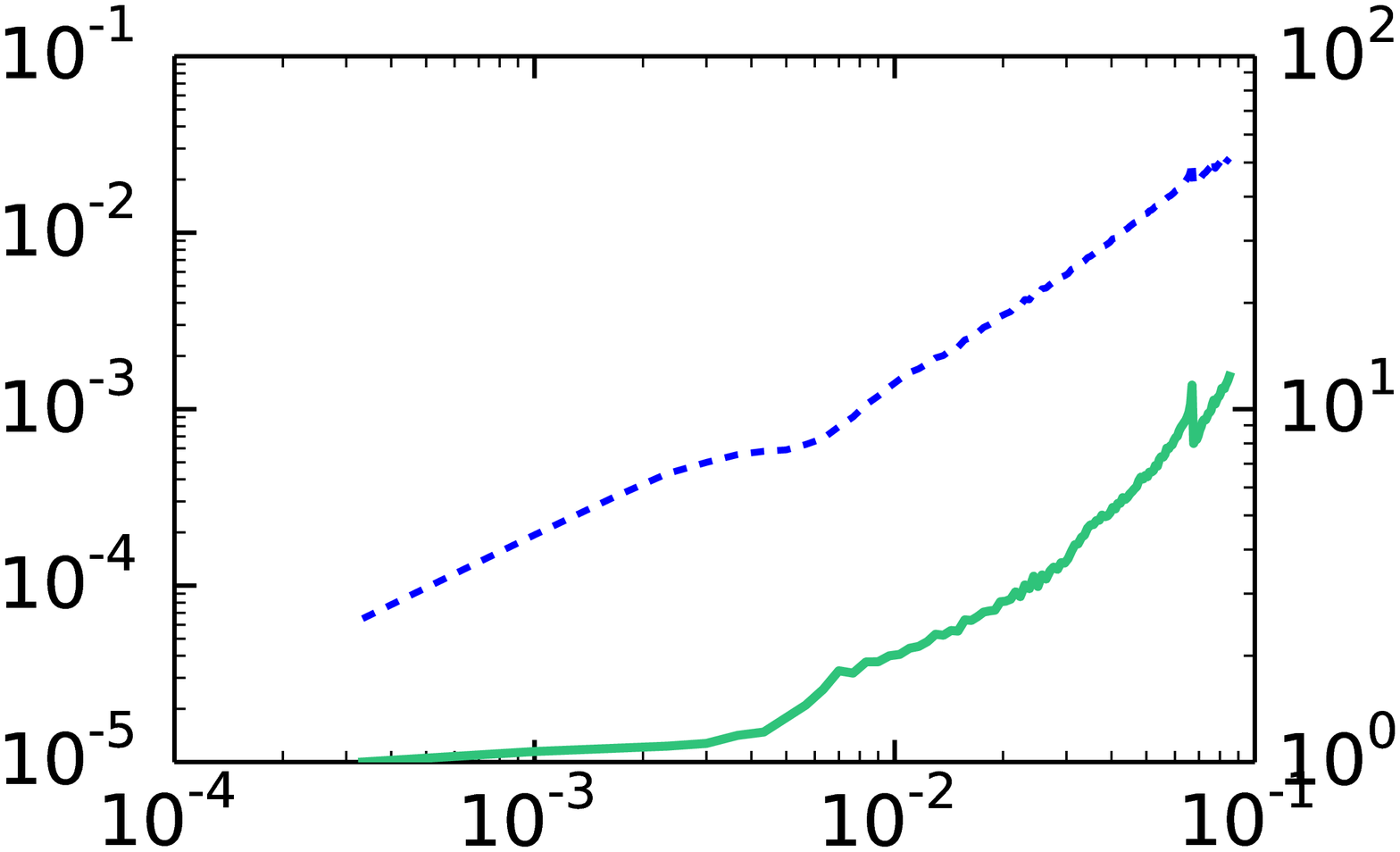}
		\put(50,35){$\dtil_c$}
		\put(-2.5,30){\rotatebox{90}{\colorbox{white}{\textcolor{myblue2}{$\Ftil$}}}}
		\put(47,0){\rotatebox{0}{\colorbox{white}{$\dtil$}}}
		\put(93.5,30){\rotatebox{90}{\colorbox{white}{\textcolor{mygreen2}{$N_s$}}}}
	\end{overpic}}
\caption{(Color online) The number $N_{s}$ of atoms in
contact (solid line) and the dimensionless impact force $\Ftil$ (dashed line) for faceted
nanoparticles that undergo facet contact
in (a) or edge contact in (b). Both nanoparticles collide at $\vcol=\unit[41]{m/s}$.}
\label{fig:FN}
\end{figure}

The kink in the number where contact area growth ends, coincides with
the transition of the force behavior at the critical overlap
\({\widetilde{\delta}}_{c}\). For the small nanoparticle of
\(R = \unit[2.5]{nm}\) the kink is identified at
\(\widetilde{\delta} \sim 5 \times 10^{- 3}\) in Fig.~4(a). Thus, we
conclude that the critical overlap \({\widetilde{\delta}}_{c}\) is the
onset of completion of contact of the first atomic layer of the facet,
and that no further growth of the area occurs at
\(\widetilde{\delta} > {\widetilde{\delta}}_{c}\).

The constant contact areas presented at
\(\widetilde{\delta} > {\widetilde{\delta}}_{c}\) differ from the
results for nanoparticles having a facet-free surface in Refs.~\cite{vergeles1997,yi2005,valentini2007,jung2012} and in our
amorphous nanoparticles and edge contacts. The nanoparticles reported in
the articles are crystalline nanoparticles with nearly spherical shapes.
They have relatively smooth surfaces, although atomic asperity exists.
Contact surface areas expand as the nanoparticles get compressed and
follow either the Hertz contact theory or the Johnson-Kendall-Roberts
theory \cite{johnson1971}, in which surface adhesion is incorporated.

At low compression \(\widetilde{\delta} < {\widetilde{\delta}}_{c}\),
where partial contact of the facet occurs, the rising rate of the
contact force with increasing overlap is higher. The number \(N_{s}\) of
contacting atoms in this compression domain rises steadily and quickly
in a very narrow range of overlap. The dynamic response triggered by the
impact load then becomes sharper than that at the higher compression,
\(\widetilde{\delta} > {\widetilde{\delta}}_{c}\).

In the edge contact case, on the other hand, Fig.~4(b), the
very low \(N_{s}\) at \(\widetilde{\delta} < \dtil_c\) shows that only one atom is
involved in contact. The \(N_{s}\) then increases gradually, which commences in the vicinity of the critical overlap. This means that the nanoparticles are in contact via only a
single atom at the very early stage of contact, and for a longer period
of contact compared to the facet contact case. The slower compression
process happening in a very limited area on the surface is presumably a
primary reason that the contact force at
\(\widetilde{\delta} < {\widetilde{\delta}}_{c}\) increases more slowly
by comparison to that at $\dtil > \dtil_c$.

\subsection{Hertzian or
non-Hertzian?}\label{b.-hertzian-or-non-hertzian}

The failure of the Hertz contact law for facet contact in our
crystalline nanoparticles at high compression seems to be attributable
to the occurrence of an impact on a large contact area. Since the forces
for edge contact and amorphous surface contact shown earlier recover at
least the slope of the Hertzian contact force, for sufficient
compression. Two past papers may give us another perspective on
repulsive nanoparticles in contact.

In an MD-based nanoindentation study \cite{bian2013}, the contact force of
a small copper nanoparticle of radius \(R = \unit[10]{nm}\) under uniaxial
compression in its \{001\} facets by two opposing rigid plates behaves
according to the Hertzian model. The copper atoms in the simulation were
modeled by the Embedded Atom Model (EAM) \cite{baskes1992} and the pure
repulsion between the plates and the copper nanoparticle adopted a
harmonic potential.

Moreover, in Ref.~\cite{kuninaka2009a} a facet collision for two LJ nanoparticles with a
pure repulsion acting between the nanoparticles described by a soft
potential \(4\epsilon(\sigma/r_{\text{ij}})^{12}\) with a cutoff
\(\unit[2.5]{\sigma}\), which is the repulsive term in the 12-6 LJ potential, was
reported. Its force on the contact surface turned out to be consistent
with Hertzian contact theory.

In both papers, the (100) facet on fcc nanoparticles was chosen for
contact. The results are inconsistent with our results for facet
contact. This discrepancy perhaps arises from the choices in surface
interactions, the softer interactions employed in the above papers and
the stiffer interaction, i.e. the WCA potential, used in our simulation.
Further investigation using different potentials is needed to identify the disagreement.

\section{Conclusions}\label{conclusions}

We have presented the quasi-elastic interactions of two approximate
spherical nanoparticles that undergo a head-on collision obtained by
means of non-equilibrium MD simulation. The pure
repulsive impact force between two nanoparticles in contact was achieved
by adopting the Weeks-Chandler-Andersen potential. To study the effect
of atomic scale surface roughness and structure on the impact force,
monocrystalline nanoparticles produced from a face-centered cubic crystal
and amorphous nanoparticles were prepared. Crystalline nanoparticles
possess crystal facets, steps, and sharp edges on their surfaces, while
amorphous nanoparticles have comparatively smooth surfaces, although
some atomic roughness still remains. We have compared our numerical
results with macroscopic counterparts described by Hertzian contact
theory.

For monocrystalline nanoparticles, two different contact types, a facet
contact and an edge contact, arising from the surface geometry of the
faceted nanoparticles were taken into account. The impact on the facets
causes dynamic properties to deviate considerably from predictions of
the Hertz contact theory. The collision on facets of both particles
reveals a non-Hertzian contact force \(F_{N} \propto \delta^{n}\), where
\(\delta\) denotes the overlap and \(n < 3/2\) when the nanoparticles
are sufficiently compressed. The mutual contact surface area of the
colliding nanoparticles does not expand in the quasi-elastic collision
regime.

In contrast, during the edge collision, the mutual contact surface area
expands, as opposed to the facet collision case. The impact force for
the edge collision shows excellent quantitative agreement with the
Hertzian contact force when compression is sufficiently high,
\(\delta/R > 10^{- 2}\).

For a system in which orientation of the nanoparticles cannot be
controlled, the possibility of precise facet contact is not high.
Overall collisional behavior of the nanoparticles is thought to be
dominated by edge contact; hence, Hertzian contact theory is valid in
this regard. However, a crystallographic orientation-controlled system, for instance the
oriented attachment technique \cite{halder2007}, may have to consider the non-Hertzian
interactions we have demonstrated, if nanoparticles are compelled to
impact on facets together.

In addition to the crystalline nanoparticle case, collisions of
amorphous nanoparticles, which have a rather spherical shape, were
simulated in the same manner to examine the influence of facet-free
surface structure. The impact force of amorphous nanoparticles agrees
qualitatively, \(F_{N} \propto \delta^{n}\), where \(n \approx 1.5\).
However, a dynamic effect on the impact force is observed.

In both crystalline and amorphous nanoparticles, a departure from the
Hertzian contact force is commonly seen when nanoparticles are weakly
compressed, \(\delta/R < 10^{- 2}\). In this compression regime, we
found that only a few atoms on each contact surface are involved in
contact and surface discreteness influences the impact force.

\section*{Acknowledgments}
We thank the U.S. Army Research Office for partial support of the present research.

\section*{References}
\bibliographystyle{elsarticle-num}
\bibliography{The_effect_of_surface_geometry_on_collisions_between_nanoparticles}

\begin{thebibliography}{10}
\expandafter\ifx\csname url\endcsname\relax
  \def\url#1{\texttt{#1}}\fi
\expandafter\ifx\csname urlprefix\endcsname\relax\def\urlprefix{URL }\fi
\expandafter\ifx\csname href\endcsname\relax
  \def\href#1#2{#2} \def\path#1{#1}\fi

\bibitem{luan2005}
B.~Luan, M.~O. Robbins, The breakdown of continuum models for mechanical
  contacts, Nature 435~(7044) (2005) 929--932.

\bibitem{hertz1881}
H.~Hertz, On the contact of elastic solids, J. Reine Angew. Math. 92 (1881)
  156--171.

\bibitem{pastewka2016}
L.~Pastewka, M.~O. Robbins, Contact area of rough spheres: Large scale
  simulations and simple scaling laws, Applied Phys. Lett. 108~(22).

\bibitem{barnard2009}
A.~S. Barnard, N.~P. Young, A.~I. Kirkland, M.~A. Van~Huis, H.~Xu, Nanogold: a
  quantitative phase map, ACS Nano 3~(6) (2009) 1431--1436.

\bibitem{dominik1996}
C.~Dominik, A.~G. G.~M. Tielens, Resistance to sliding on atomic scales in the
  adhesive contact of two elastic spheres, Philos. Mag. A 73~(5) (1996)
  1279--1302.

\bibitem{awasthi2007}
A.~Awasthi, S.~C. Hendy, P.~Zoontjens, S.~A. Brown, F.~Natali, Molecular
  dynamics simulations of reflection and adhesion behavior in lennard-jones
  cluster deposition, Phys. Rev. B 76~(11)  115437.

\bibitem{luding1999}
S.~Luding, H.~J. Herrmann, Cluster-growth in freely cooling granular media,
  Chaos 9~(3).

\bibitem{xu2016}
J.~Xu, B.~Zheng, Y.~Liu, Solitary wave in one-dimensional buckyball system at
  nanoscale, Scientific Reports 6 (2016) 21052 EP --.

\bibitem{nesterenko1983}
V.~F. Nesterenko, Propagation of nonlinear compression pulses in granular
  media, J. Appl. Mech. Tech. Phys. 24~(5) (1983) 733--43.

\bibitem{nesterenko2001}
V.~Nesterenko, Dynamics of Heterogeneous Materials, Springer, 2001.

\bibitem{sen2008}
S.~Sen, J.~Hong, J.~Bang, E.~Avalos, R.~Doney, Solitary waves in the granular
  chain, Physics Reports 462~(2) (2008) 21 -- 66.

\bibitem{kuninaka2009a}
H.~Kuninaka, H.~Hayakawa, Super-elastic collisions in a thermally activated
  system, Prog. Theor. Phys. Suppl. 178 (2009) 157--162.

\bibitem{zeng2010}
Q.~Zeng, A.~Yu, G.~Lu, Evaluation of interaction forces between nanoparticles
  by molecular dynamics simulation, Ind. Eng. Chem. Res. 49~(24) (2010)
  12793--12797.

\bibitem{kuninaka2009b}
H.~Kuninaka, H.~Hayakawa, Simulation of cohesive head-on collisions of
  thermally activated nanoclusters, Phys. Rev. E 79 (2009) 031309.

\bibitem{han2010}
L.~B. Han, Q.~An, S.~N. Luo, W.~A. Goddard~III, Ultra-elastic and inelastic
  impact of cu nanoparticles, Mater. Lett. 64~(20) (2010) 2230--2232.

\bibitem{takato2014}
Y.~Takato, S.~Sen, J.~B. Lechman, Strong plastic deformation and softening of
  fast colliding nanoparticles, Phys. Rev. E 89~(3) (2014) 033308.

\bibitem{takato2015}
Y.~Takato, M.~E. Benson, S.~Sen, Rich collision dynamics of soft and sticky
  crystalline nanoparticles: Numerical experiments, Phys. Rev. E 92 (2015)
  032403.

\bibitem{thornton1998}
C.~Thornton, Z.~Ning, A theoretical model for the stick/bounce behaviour of
  adhesive, elastic-plastic spheres, Powder technol. 99~(2) (1998) 154--162.

\bibitem{awasthi2006}
A.~Awasthi, S.~C. Hendy, P.~Zoontjens, S.~A. Brown, Reentrant adhesion behavior
  in nanocluster deposition, Phys. Rev. Lett. 97~(18) (2006) 186103.

\bibitem{jung2010}
S.-c. Jung, D.~Suh, W.-s. Yoon, Molecular dynamics simulation on the energy
  exchanges and adhesion probability of a nano-sized particle colliding with a
  weakly attractive static surface, J. Aerosol Sci. 41~(8) (2010) 745--759.

\bibitem{weeks1971}
J.~D. Weeks, D.~Chandler, H.~C. Andersen, Role of repulsive forces in
  determining the equilibrium structure of simple liquids, J. Chem. Phys.
  54~(12) (1971) 5237--5247.

\bibitem{johnson1987}
K.~Johnson, Contact Mechanics, Cambridge Univiersity Press, 1987.

\bibitem{sun2011}
D.~Sun, C.~Daraio, S.~Sen, Nonlinear repulsive force between two solids with
  axial symmetry, Phys. Rev. E 83 (2011) 066605.

\bibitem{kristensen1976}
W.~D. Kristensen, Computer-simulated amorphous structures (i). quenching of a
  lennard-jones model system, J. Non-Cryst. Solids 21~(3) (1976) 303--318.

\bibitem{nose1985}
S.~Nos{\'e}, F.~Yonezawa, Isobaric-isothermal molecular dynamics study on the
  glass transition of a lennard-jones system, Solid State Commun. 56~(12)
  (1985) 1005--1008.

\bibitem{rahman1976}
A.~Rahman, M.~Mandell, J.~McTague, Molecular dynamics study of an amorphous
  lennard-jones system at low temperature, J. Chem. Phys 64~(4) (1976)
  1564--1568.

\bibitem{plimpton1995}
S.~Plimpton, Fast parallel algorithms for short-range molecular dynamics, J.
  Comput. Phys. 117~(1) (1995) 1 -- 19.

\bibitem{humphrey1996}
W.~Humphrey, A.~Dalke, K.~Schulten, Vmd - visual molecular dynamics, J. Molec.
  Graphics 14 (1996) 33 -- 38.

\bibitem{kim2010}
S.~Kim, Elastic behavior of spherical nanodroplets in head-on collisions, J.
  Korean Phys. Soc. 56 (2010) 969--972.

\bibitem{vergeles1997}
M.~Vergeles, A.~Maritan, J.~Koplik, J.~R. Banavar, Adhesion of solids, Phys.
  Rev. E 56~(3) (1997) 2626.

\bibitem{yi2005}
M.~Y. Yi, D.~S. Kim, J.~W. Lee, J.~Koplik, Molecular dynamics (md) simulation
  on the collision of a nano-sized particle onto another nano-sized particle
  adhered on a flat substrate, J. Aerosol Sci. 36~(12) (2005) 1427--1443.

\bibitem{valentini2007}
P.~Valentini, T.~Dumitric{\u{a}}, Microscopic theory for nanoparticle-surface
  collisions in crystalline silicon, Phys. Rev. B 75~(22) (2007) 224106.

\bibitem{jung2012}
S.-C. Jung, J.-G. Bang, W.-s. Yoon, Applicability of the macro-scale elastic
  contact theories for the prediction of nano-scaled particle collision with a
  rigid flat surface under non-adhesive and weakly-adhesive conditions, J.
  Aerosol Sci. 50 (2012) 26--37.

\bibitem{johnson1971}
K.~Johnson, K.~Kendall, A.~Roberts, Surface energy and the contact of elastic
  solids, Proc. R. Soc. Lond. A, Math. Phys. Sci. 324~(1558) (1971) 301--313.

\bibitem{bian2013}
J.-J. Bian, G.-F. Wang, Atomistic deformation mechanisms in copper
  nanoparticles, Journal of Computational and Theoretical Nanoscience 10~(9)
  (2013) 2299--2303.

\bibitem{baskes1992}
M.~Baskes, Modified embedded-atom potentials for cubic materials and
  impurities, Phys. Rev. B 46~(5) (1992) 2727.

\bibitem{halder2007}
A.~Halder, N.~Ravishankar, Ultrafine single-crystalline gold nanowire arrays by
  oriented attachment, Adv. Mater. 19~(14) (2007) 1854--1858.

\end{thebibliography}

\end{document}